\documentclass[twocolumn,showkeys]{aastex631}

\usepackage{amsmath}   % \dfrac, cases; aastex631 does not load it

\usepackage{subcaption}
\usepackage{booktabs}

\makeatletter
\expandafter\let\csname longtable*\endcsname\@undefined
\expandafter\let\csname endlongtable*\endcsname\@undefined
\makeatother

\AtBeginDocument{}

\newcommand{\tablefoot}[1]{%
  \par\vspace{2pt}%
  \begin{minipage}{\linewidth}\footnotesize\textsc{Note}---#1\end{minipage}}
\usepackage{xspace}
\newcommand{\numax}{\ensuremath{\nu_{\max}}\xspace}
\newcommand{\dnu}{\ensuremath{\Delta\nu}\xspace}

\begin{document}

\title{ROSE (Red-giant Oscillations Spectra Estimator): A Modular Machine-Learning Framework for Automated Asteroseismic Characterisation. I. $\nu_{\max}$ and $\Delta\nu$ from \emph{TESS}}

\shorttitle{ROSE I. $\nu_{\max}$ and $\Delta\nu$ from \emph{TESS} Red Giants}
\shortauthors{Ghanghas et al.}

\correspondingauthor{Nipun Ghanghas}
\email{nipun@tifr.res.in}

\author[0009-0004-4477-3870]{Nipun Ghanghas}
\affiliation{Department of Astronomy and Astrophysics, Tata Institute of
Fundamental Research, Homi Bhabha Road, Mumbai 400005, India}

\author[0000-0002-6812-4443]{Dinil B. Palakkatharappil}
\affiliation{Universit\'e Paris-Saclay, Universit\'e Paris Cit\'e, CEA, CNRS,
AIM, 91191, Gif-sur-Yvette, France}

\author[0000-0002-8854-3776]{Rafael A. Garc\'ia}
\affiliation{Universit\'e Paris-Saclay, Universit\'e Paris Cit\'e, CEA, CNRS,
AIM, 91191, Gif-sur-Yvette, France}

\author[0000-0003-2896-1471]{Shravan Hanasoge}
\affiliation{Department of Astronomy and Astrophysics, Tata Institute of
Fundamental Research, Homi Bhabha Road, Mumbai 400005, India}
\affiliation{Center for Space Science, NYUAD Institute, New York University Abu
Dhabi, PO Box 129188, Abu Dhabi, UAE}

\begin{abstract}
Space-based photometry from \emph{Kepler} and \emph{TESS} has delivered oscillation spectra for hundreds of thousands of red giants, and PLATO and the Roman Galactic Bulge Time-Domain Survey will add more. We present ROSE, a modular machine-learning framework for automated asteroseismic characterisation of red giants, and apply two of its modules to the frequency of maximum oscillation power, $\nu_{\max}$, and the large frequency separation, $\Delta\nu$. Each module is a neural network bundled with its own preprocessing transform and output binning, so that each parameter is measured in the representation that displays it most effectively: $\nu_{\max}$ from the power-density spectrum, and $\Delta\nu$ from the power spectrum of the power spectrum taken over a window centred on $\nu_{\max}$. Both are trained only on synthetic spectra generated at a one-year baseline and both return probability distributions rather than point estimates. This enables us to retrieve asymmetric uncertainty and a reliability index. Validating against held-out synthetic spectra, \emph{Kepler} spectra at the training resolution and \emph{TESS} continuous-viewing-zone giants, we obtain robust dispersions of $1.4$, $3.3$ and $4.4\%$ in $\nu_{\max}$ and $0.35$, $0.61$ and $1.3\%$ in $\Delta\nu$, with the $\nu_{\max}$ uncertainties close to their nominal coverage. The fully automated pipeline was applied to $274\,721$ \emph{TESS} red-giant candidates observed for at least ten sectors, the framework returning reliable $\nu_{\max}$ for $56\,224$ stars and reliable $\Delta\nu$ for $39\,939$, taking $8$\,ms per star. The sample follows $\Delta\nu = 0.278\,\nu_{\max}^{0.757}$ with a scatter of $6.2\%$, and the red-clump phase stands out as a distinct overdensity, although neither module is provided an evolutionary label or asked to produce one.
\end{abstract}


\section{Introduction}\label{sec:intro}
Red giants are among the most useful stars in the galaxy for asteroseismology. Their solar-like oscillations have large amplitudes and periods of hours to days, so they remain detectable at kiloparsec distances with photometry of modest precision, and two global parameters derived from the oscillation spectrum carry a great deal of information about the star. The frequency of maximum oscillation power, $\nu_{\max}$, scales with the acoustic cut-off frequency and therefore with surface gravity. The large frequency separation, $\Delta\nu$, measures the mean density through the spacing of consecutive pressure modes. Combined with an effective temperature, the two give a radius and a mass through the asteroseismic scaling relations \citep{Huber_2011, Garcia_2019_review}, and with stellar models an age \citep{davies2016AsteroseismologyRedGiantsa}. Ages obtained this way are precise enough to be used as a dating tool for the galactic disc \citep{Anders_2017, Sharma_2019, Miglio_2021}. For this, we need large and uniformly analysed samples.

\emph{Kepler} \citep{Kepler_mission} established that such samples were possible. Four years of nearly uninterrupted photometry yielded $\nu_{\max}$ and $\Delta\nu$ for some sixteen thousand red giants \citep{Yu_2018}, a catalogue still used as the reference against which many later works are compared. \emph{TESS} \citep{Tess_mission_ricker_2015} changed the scale of the problem rather than the quality of the data. Oscillations have been detected and $\nu_{\max}$ measured for about $1.6\times10^{5}$ giants across most of the sky from single sectors of full-frame-image photometry \citep{quick_look_Hon_el_al}, and over continuous viewing zones, where the baseline can reach several years, both parameters are now available for tens of thousands of stars \citep{Mackereth_2021MNRAS.502.1947M, Hon_2022AJ....164..135H, Sreenivas_2026}. PLATO \citep{plato_mission_2024} and the Roman Galactic Bulge Time-Domain Survey \citep{Spergel_2015_Roman, Huber_2023_Roman} will add large datasets as well, the latter mostly of bulge giants that no previous mission has been able to study.

At this scale, the analysis itself becomes the limiting step, and the two parameters are not equally difficult. Locating a broad power excess is easier: the classical pipelines developed for \emph{Kepler}, among them SYD \citep{Huber_2009_SYD}, COR \citep{Mosser_2009_COR}, OCT \citep{Hekker_2010_BHM}, A2Z \citep{Mathur_2010_A2Z}, and CAN \citep{Kallinger_2010_CAN}, agree on $\nu_{\max}$ to within a few per cent even on shortened time series. Measuring $\Delta\nu$ is harder, because it requires the individual modes to be resolved and the comb they form to be identified rather than merely detected. With one or two \emph{TESS} sectors, $\Delta\nu$ is recovered for only $14$--$26\%$ of \emph{Kepler}-field giants that \emph{TESS} re-observed \citep{Stello_2022MNRAS.512.1677S}. Even with the full continuous-viewing-zone baseline, the catalogue of \citet{Sreenivas_2026} recovers $\Delta\nu$ of $58\%$ of its targets and flags the rest, and several \emph{TESS} studies have needed case-by-case human inspection to reach that point \citep{Hon_2022AJ....164..135H}. Visual vetting works, but it does not scale to millions of light curves that current and future surveys provide, and it leaves the selection function of the resulting catalogue difficult to model.

Machine learning can supply the throughput. Convolutional networks classify evolutionary state \citep{Hon2017}, detect oscillations and estimate $\nu_{\max}$ from spectra treated as images \citep{Hon_2018, hon_kepler_giant_detections, quick_look_Hon_el_al}, and more recently, infer $\Delta\nu$ together with the mixed-mode parameters $\Delta\Pi_1$ and $q$ \citep{dhanpal2022, dhanpal2023, ghanghas2026periodspacingsglobalseismic}. Training on simulated spectra - as the latter group of studies does - removes the dependence on a reference catalogue and allows the training distribution to be chosen rather than inherited. Recasting the inference as a classification over parameter bins, rather than as a regression to a single number, turns the output into a probability distribution from which an uncertainty follows directly \citep{bishop1994_mixture_density_networks, Rothe_2015, Stewart_2022arXiv221105641S}.

Speed, however, is not the only thing that matters, and there are additional considerations to take into account here. A network learns the regularities its training data contain, and $\nu_{\max}$ and $\Delta\nu$ are tightly correlated in observed red giants. A model asked for both from a single input representation can therefore exploit that correlation instead of measuring each quantity independently, and the risk rises where the data are of poorest quality or the mode pattern hard to read. A catalogue produced that way will reproduce the expected relation, which is usually the first thing tested. But $\Delta\nu$ obtained in that manner carries little information that $\nu_{\max}$ did not already carry, and later work that treats the two as independent measurements, which includes essentially every mass and radius determination, would inherit the scaling relation. While $\Delta\nu$ vetting can be performed in a fast automated way \citep{reyes2022VettingAsteroseismicDn}, this can also exclude many stars for which $\Delta\nu$ is actually measurable.

The way to avoid this is not to train harder but to choose the training set and the input representation with the problem in mind, and the same choices lend precision as well as robustness. We take two measures in this work to circumvent these issues. The first concerns the synthetic spectra, in which $\nu_{\max}$ is drawn with a spread of $30\%$ about the value implied by $\Delta\nu$, which is several times higher than the typical spread observed in real red giants \citep{Stello_2009,Yu_2018}. The relation then carries much weaker constraining information than observations typically support, and a network cannot lean on it during training. The second is that $\Delta\nu$ is inferred by a module of its own, which works not on the power spectrum but on the power spectrum of the power spectrum (PSPS) taken in a window cropped around the oscillation envelope, where a regular comb of modes appears as a peak at the comb spacing. Classical methods have long exploited this representation \citep{Mosser_2009_COR, Hekker_2010_BHM}, and what it offers a network is a more direct view of the quantity being measured rather than an indirect one. We find this to be the better route in both regimes, at high signal-to-noise as well as low. The module uses $\nu_{\max}$ only to decide which part of the spectrum to crop, and not as a predictor of the answer. Thus, the $\Delta\nu$--$\nu_{\max}$ relation stays outside the method entirely and remains available afterwards as a check on the predictions.

This separation and flexibility is the organising idea of the framework we present here, a modular machine-learning framework for the asteroseismic characterisation of red giants (ROSE). A module is one network trained to infer one parameter, bundled with the preprocessing transform that defines its input and the binning scheme that defines its output. The configuration of both is stored inside the saved model file, so the data loaders prepare the input in the expected form automatically. Modules can be retrained, replaced, or added without disturbing the rest of the pipeline, and each can be given the representation best suited to the parameter it measures. Because every module returns a probability distribution rather than a point estimate, each prediction carries an asymmetric uncertainty together with diagnostics of the shape of that distribution, any multi-modalities if present, and these are what we use to decide which predictions the framework should report.

We train two modules on a set of $\sim 2$ million synthetic spectra generated at a one-year baseline, and validate them on three samples of decreasing control: held-out synthetic spectra, \emph{Kepler} red giants resampled to the same one-year resolution and compared with \citet{Yu_2018}, and \emph{TESS} continuous-viewing-zone giants compared with \citet{Sreenivas_2026}. The residual dispersion is $1.4\%$, $3.3\%$ and $4.4\%$ in $\nu_{\max}$ and $0.35\%$, $0.61\%$ and $1.3\%$ in $\Delta\nu$, respectively. The $\nu_{\max}$ uncertainties are close to their nominal coverage throughout; the $\Delta\nu$ intervals are conservative on the first two samples. We then apply the framework to $274\,721$ red-giant candidates observed by \emph{TESS} in at least ten sectors, obtaining reliable $\nu_{\max}$ for $56\,224$ and reliable $\Delta\nu$ for $39\,939$, at a cost of a few milliseconds of GPU time per star, with no visual inspection at any stage. The red-clump overdensity in the resulting $\nu_{\max}^{0.75}/ \Delta\nu - \nu_{\max}$ plane is prominently visible, without any evolutionary information having been supplied to either module, which confirms that red clump stars are strongly enriched in the \emph{TESS} sample and that the networks measure both parameters with good precision for them.

The paper is organised as follows. Section~\ref{sec:data} describes the \emph{TESS} and \emph{Kepler} samples and the reference catalogues. Section~\ref{sec:methods} sets out the framework: the synthetic training set, the two modules and their input representations, treatment of frequency resolution during inference, uncertainty and reliability measures, and the selection built from them. Section~\ref{sec:results} presents the three validation comparisons and the application to the full \emph{TESS} sample, and Section~\ref{sec:conclusion} summarises our conclusions.

\section{Data}
\label{sec:data}

\subsection{The TESS sample}
\label{sec:data-tess}

Candidate red giants were identified from \emph{Gaia} \citep{Gaia_DR2_2018A&A...616A...1G} by combining an XGBoost classification with the position of each star in the Kiel and colour--magnitude diagrams, and the resulting list was cross-matched against the \emph{TESS} input catalogue \citep{Tess_mission_ricker_2015}. The selection is photometric and intentionally broad. It is meant to enclose the giant branch rather than to return a list of confirmed oscillators, and the yields quoted in Sect.~\ref{sec:res-catalogue} should be read with that in mind.

Light-curves were built from the detrended full-frame-image photometry of the MIT Quick-Look Pipeline \citep[QLP;][]{Huang_2020_QLP_I, Huang_2020_QLP_II}, with custom processing applied on top by PyTADACS (Python \emph{TESS} Automated Data Analysis and Correction Software; \citealp{garcia2024measuringrotationperiods,garcia2024massivecharacterisation}; Palakkatharappil et al., in prep.), the \emph{TESS} descendant of the KADACS pipeline written for \emph{Kepler} \citep{Garcia_2011_KADACS, garcia2014ImpactGapsKepler, Pires_2015}. Gaps longer than three sectors are removed by shifting the timestamps of the points after the gap to one cadence after the last point preceding it. This minimises the effect of the window function in the frequency domain while reducing the correlation between points; see Appendix~C of \citet{Beck_2026} for more details. The sectors are then stitched together by matching their median fluxes, and outliers are rejected with a $5\sigma$ clip. The light curve is then placed on a regular $30$-minute cadence by nearest-neighbour resampling under the slotting principle described by \citet{garcia2014ImpactGapsKepler}. This fixes the Nyquist frequency at $277.78\,\mu$Hz, which is the upper edge of the frequency grid used throughout this work (Sect.~\ref{sec:training-set}), and it puts the sampling of the \emph{TESS} light curves on the same footing as \emph{Kepler} long-cadence photometry.

Slow instrumental and astrophysical trends are removed by subtracting a triangular smoothing of the light curve with a full width of five days. This suppresses variability on timescales longer than 5 days, and therefore attenuates the power spectrum below a characteristic frequency of $1/5\,\mathrm{d} \approx 2.3\,\mu$Hz. The attenuation lies well below the oscillation envelope of all but the most evolved giants in our range, but it is not negligible at the low-frequency end, and we return to its consequences in Sect.~\ref{sec:res-tess}. Power spectra are computed using the Lomb--Scargle periodogram \citep{Lomb_1976, Scargle_1982}.

We retain stars with photometry in at least ten sectors in total. The sectors need not be consecutive, since the gap treatment described above is what allows a discontinuous record to be analysed as though it were continuous. The criterion we need the sample to meet is that it must contain enough accumulated signal to approach the one-year baseline of the training set (Sect.~\ref{sec:training-set}). This gives a sample of $274\,721$ stars. Each spectrum is then placed on the fixed one-year frequency grid of Sect.~\ref{sec:resolution} before being passed to the modules, since the networks require inputs of uniform resolution and the available baseline varies from star to star.

\subsection{The Kepler sample}
\label{sec:data-kepler}

To test the framework on real stars while keeping the reference under control, we use the sample of $\sim21\,000$ \emph{Kepler} red giants assembled by \citet{dhanpal2022}. Long-cadence light curves for all available quarters are retrieved with \texttt{lightkurve} \citep{Lightkurve_2018}, stitched, cleared of non-finite values and outliers, and normalised to parts per million. The Lomb--Scargle periodogram is then evaluated directly on the $8760$-point grid spanning $0.03$--$277.78\,\mu$Hz, so that the spectrum handed to the modules is sampled at exactly the frequency spacing they were trained on.

This construction is the analogue, at a one-year baseline, of the \emph{Kepler}-as-K2 test set of \citet{ghanghas2026periodspacingsglobalseismic}. The frequency grid and the $\chi^2_2$ statistics of each spectral bin match those of a one-year observation, but the underlying mode profiles retain the sharpness conferred by the full four-year window rather than being broadened to the one-year resolution. The test is therefore sampling-matched rather than a full degradation to one-year data, and for features narrower than the one-year grid spacing it is mildly optimistic. The oscillation envelope and the pressure modes are however well resolved at a one-year baseline, so the effect is practically negligible for $\nu_{\max}$ and $\Delta\nu$. The reference values are taken from the full four-year light curves and are more precise than a one-year resolution can afford, so the residual scatter remains dominated by framework error.

\subsection{Reference catalogues}
\label{sec:data-refs}

For \emph{Kepler}, we compare against \citet{Yu_2018}, who measured $\nu_{\max}$ and $\Delta\nu$ for $16\,095$ red giants from the full four-year light curves. The cross-match with our sample yields $15\,799$ stars.

For \emph{TESS} we compare against \citet{Sreenivas_2026}, who measured the same two quantities for $19\,151$ red giants in the continuous viewing zones. That analysis uses the full available baseline rather than a fixed one, with a median of $33$ sectors spanning $690$ days; $28\%$ of their targets exceed two years and the longest samples reach $3.7$ years. The cross-match with our sample yields $17\,203$ stars. This reference is therefore more precise than the one-year spectra we feed to the modules, but by a smaller margin than in the \emph{Kepler} case: a median baseline of $690$ days against one year, rather than four years against one. There is a second difference that matters more. Both analyses are performed using data from the same instrument, so they share the systematics that \emph{TESS} photometry carries, and the comparison tests our framework against an independent pipeline rather than against a different mission. We also use their $\Delta\nu$ reliability flag, \texttt{f\_dnu}, as described in Sect.~\ref{sec:res-tess}.

\section{Methods}
\label{sec:methods}

\subsection{Overview of the framework}
\label{sec:overview}

The framework is organised as a set of independent \emph{modules}, each trained to infer a single seismic parameter. A module bundles: a preprocessing transform, which defines its input, a neural network, and a binning scheme, which defines its output. Inferences from all modules share a similar interface. They take in a power spectral density (PSD) on an arbitrary frequency grid, along with any auxiliary quantities they require, and return a probability distribution over that parameter. Nothing else is shared, so a module can be retrained, replaced or chained with others without affecting the rest of the pipeline.

This differs from the design adopted in \citet{ghanghas2026periodspacingsglobalseismic}, where a single network predicted $\nu_{\max}$, $\Delta\nu$, $\Delta\Pi_1$ and $q$ jointly from one input and the four outputs were coupled through shared loss weights. Joint training was advantageous there because the down-weighted outputs acted as auxiliary tasks that suppressed catastrophic outliers. However, this is rigid: the input representation that is optimal for locating a broad power excess is not optimal for measuring a comb spacing, and a single input forces a compromise between them. Separating the problem allows each module to be given the representation best suited to it, at the cost of having to propagate one module's output into the next.

In this work, we use two modules. The first, which we refer to as the $\nu_{\max}$ module, infers the frequency of maximum oscillation power from the full-range PSD. The second, the $\Delta\nu$ module, infers the large frequency separation from the PSPS computed over a window centred on $\nu_{\max}$; it therefore depends on the $\nu_{\max}$ module, which supplies that centre. Both are trained exclusively on synthetic spectra generated at a fixed one-year baseline (Sect.~\ref{sec:training-set}), and similar to \citet{dhanpal2022, dhanpal2023, ghanghas2026periodspacingsglobalseismic}. Both recast the inference as a classification over discrete bins so that the output is a probability distribution rather than a point estimate (Sect.~\ref{sec:uncertainties}).

Because a module is only meaningful together with its preprocessing transform, the complete training configuration -- the frequency grid and its resolution, the normalisations applied, the window and smoothing parameters, and the output bin edges and centres -- is serialised and stored inside the saved model file itself. During inference, the transform is reconstructed from that record rather than restated by the user, which removes the most common failure mode of this kind of pipeline: a spectrum prepared in a way that the network did not encounter during training. Sect.~\ref{sec:resolution} describes one such failure that this mechanism prevents.

\subsection{Training set}
\label{sec:training-set}
Both modules are trained on synthetic power spectra, for the reason set out in \citet{ghanghas2026periodspacingsglobalseismic} and section \ref{sec:intro}: the number of red giants with reliable reference measurements is small compared with the volume of data a network of this size requires, and training on catalogue values would inject the selection function and the systematics of the catalogue into the network. Synthetic spectra allow the parameters to be sampled independently or weakly correlated with one another, so that the network cannot recover one parameter as a by-product of a learned correlation with another.

Spectra are generated using the Spectra Simulator framework \citep{othman_benomar_2023_spectra_simulator}, in the implementation described in Appendix~A of \citet{ghanghas2026periodspacingsglobalseismic}. Mode frequencies follow the asymptotic description of pressure modes \citep{Mosser_2012_evolution} and the implicit mixed-mode relation \citep{Mosser_2015}; rotational splittings are included to first order through the two-zone formalism of \citet{Goupil_2013}. Heights and linewidths of dipolar mixed modes are scaled from radial-mode templates through the normalised gravity-mode inertia $\zeta(\nu)$ following \citet{Grosjean_2014A&A...572A..11G} and \citet{Mosser_2018}, with the dilution factor of \citet{Dupret_2009} applied to modes narrower than the frequency resolution so that total mode power is conserved. The background is modelled as a sum of a Harvey-like granulation component and a flat white-noise term, and the model spectrum is multiplied by a random realisation of a $\chi^2$ distribution with two degrees of freedom.

The main difference with respect to the K2 training set is the baseline, spread in \numax--\dnu and SNR range. All spectra are generated on a fixed grid of $8760$ points spanning $0.03$--$277.78\,\mu$Hz, corresponding to a frequency resolution of $\delta\nu_{\rm tr} = 0.0317\,\mu$Hz and hence to a continuous one-year window. The upper limit is not arbitrary: it is the Nyquist frequency of the $30$-minute cadence onto which the observed light curves are resampled (Sect.~\ref{sec:data-tess}), so the training grid spans exactly the range the data can carry. This resolution is a property of the training set that both modules inherit, and is enforced at inference time (Sect.~\ref{sec:resolution}).

Table~\ref{tab:parameter_ranges} lists the ranges over which the generating parameters are sampled. The parameters are drawn independently of one another, except where a scaling is stated explicitly. We again note that a very weak correlation exists in $\nu_{\max}$ and $\Delta\nu$, imposed through $\nu_{\max}/\nu_{\max,\rm scaled}$. However, it is much weaker than the observed correlation in observed red giants, so that a network attempting to exploit it is penalised rather than rewarded during training. The ranges are chosen to span the set of red giants that \emph{TESS} can observe, and the SNR range is chosen to cover the typical range of observed SNRs in the \emph{TESS} sample. The number of noise realizations per parameter set is also randomised, so that some parameter sets are represented by multiple spectra and some by only one.

\begin{table}[htbp]
\centering
\caption{Ranges of the parameters used to generate the synthetic training set. The notation follows Appendix~A of \citet{ghanghas2026periodspacingsglobalseismic}, which gives the expressions these parameters enter and the role each plays in building the model spectrum.}
\label{tab:parameter_ranges}
\setlength{\tabcolsep}{3pt}\footnotesize
\begin{tabular}{l c c}
\toprule
\textbf{Parameter} & \textbf{Range} & \textbf{Distribution} \\
\midrule
$\Delta\nu$ ($\mu$Hz)                      & $1$--$19$                  & Uniform \\
$\nu_{\max}/\nu_{\max,\rm scaled}$         & $0.7$--$1.3$      & Uniform \\
$\Delta\Pi_{1}$ (s)                        & $40$--$150$ ($\Delta\nu>9$) & Uniform \\
                                           & $40$--$500$ ($\Delta\nu<9$) & Uniform \\
$q$                                        & $0$--$0.5$        & Uniform \\
$\epsilon_{p}$                             & $0$--$1$          & Uniform \\
$\epsilon_{g}$                             & $0$--$1$          & Uniform \\
$\beta_p~(\equiv \alpha_l\,n_{\max})$      & $0.0$--$0.1$      & Uniform \\
$d_{01}$                                   & ($-0.6$ to $0.6$)/$\Delta\nu$ & Uniform \\
$d_{02}/d_{02,\rm scaled}$                 & ($0.8$ to $1.2$)/$\Delta\nu$  & Uniform \\
$d_{03}/d_{03,\rm scaled}$                 & ($0.8$ to $1.2$)/$\Delta\nu$  & Uniform \\
Core rotation ($\mu$Hz)                    & $0.005$--$2.8$    & Uniform \\
Envelope rotation ($\mu$Hz)                & $0.005$--$0.4$    & Uniform \\
Inclination $\iota$ (deg)                  & $0$--$90$         & Isotropic \\
$A_g$                                      & $0.8$--$1.2$      & Uniform \\
$B_g$                                      & $-2.2$ to $-1.8$  & Uniform \\
$C_g$                                      & $0$--$1.0$        & Uniform \\
$A_\tau$                                   & $0.8$--$1.2$      & Uniform \\
$B_\tau$                                   & $-1.0$ to $-0.9$  & Uniform \\
$C_\tau$                                   & $0$--$1.0$        & Uniform \\
$p$                                        & $1.8$--$2.4$      & Uniform \\
$N_0$                                      & $0.1$--$29\,500$           & Log-Uniform \\
Noise realisations                         & $1$--$3$          & Random \\
SNR                                        & $5$--$40$          & Uniform \\
\midrule
Frequency range ($\mu$Hz)                  & $0.03$--$277.78$           & $8760$ bins \\
Observation time (days)                    & $365.05$                   & Fixed \\
\bottomrule
\end{tabular}
\end{table}

The full dataset comprises $\sim 2$ million spectra, split into $82.5\%$ for training, $15\%$ for validation and $2.5\%$ held out for testing. Both modules are trained on the same dataset.

\subsection{The $\nu_{\max}$ module}
\label{sec:vmax-module}

\subsubsection{Input representation}

The input is the PSD over the full frequency range, resampled onto the training grid of $8760$ points spanning $0.03$--$277.78\,\mu$Hz. No background is fitted or removed and no features are pre-computed; the network operates on the spectrum directly.

The resampling depends on the input resolution. For samples in which the input spectrum is finer than the target grid, the power is averaged within each target bin, preserving the statistics of a $\chi^2_2$-distributed PSD. In contrast, when it is coarser, the power is interpolated linearly. Target bins falling outside the observed range are filled with the maximum of the first and the minimum of the last $2\,\mu$Hz of the input, so that an extrapolated low-frequency bin does not create a spurious dip in the granulation background and an extrapolated high-frequency bin does not create a spurious excess.

We trained otherwise identical variants of this module on three input representations: a linear normalised power representation, a $\log(1+P)$ transform of the normalised power, and a background-divided signal-to-noise spectrum, obtained with the sliding logarithmic median filter of Sect.~\ref{sec:psps}. The $\log(1+P)$ variant differs only marginally from the linear one and is not pursued. The signal-to-noise variant, however, behaves differently enough from the linear one that we use both, and we describe both below.

Neither representation is uniformly better; their failure modes are unrelated. The linear input retains the granulation background, and with it the low-frequency part of the spectrum that any high-pass filtering of the light curve distorts. For the \emph{TESS} spectra of Sect.~\ref{sec:data-tess}, where a five-day triangular detrending removes power below about $2.3\,\mu$Hz while the training spectra carry none, this produces a small systematic shift in $\nu_{\max}$ at the low-frequency end of the range. Dividing by a sliding logarithmic median removes any smooth multiplicative distortion of this kind automatically, so the signal-to-noise input is unaffected, as it is by other reduction-dependent instrumental effects. However, the flattening also removes the granulation background that can help locate the envelope. A spectrum containing a sharp unrelated peak, or one whose envelope is too weak to stand out, then leaves no broad feature for the network to use. The resulting $\nu_{\max}$ can then be wrong by a large factor, unlike the linear module's small systematic shift, and is reported with a narrow, confident distribution that the diagnostics of Sect.~\ref{sec:uncertainties} do not flag.

These two weak points do not coincide, so the two representations can be combined. Any small spurious peak that captures the signal-to-noise module leaves the linear module, which still sees the background, largely unaffected; the low-frequency attenuation that shifts the linear module is divided out of the signal-to-noise input. We therefore run both and require that they agree, as set out in Sect.~\ref{sec:selection}. The alternative, correcting the linear input with an explicit transfer function, was tested and works, but it requires a separately trained module for every combination of filter width and shape a reduction might use, which removes the flexibility the framework is meant to retain. A single module taking both representations as parallel inputs would likely be better than either, and we note it as the natural next step.

\subsubsection{Architecture}
Building on the final architecture adopted in \citet{ghanghas2026periodspacingsglobalseismic}, we use the same one-dimensional wide residual network as the backbone. The input passes through batch normalisation, a ReLU activation and a convolution with $16$ filters, followed by max pooling with a stride of two. The body consists of ten pre-activation residual units arranged in four stages of $[2,2,3,3]$ units with $[16,32,64,128]$ filters. Stages two to four each open with a stride-two unit, so that the number of filters doubles as the frequency axis halves. In each unit, we apply batch normalisation, ReLU, convolution, dropout and a second convolution, and add the result to its input; where the stride is greater than one, the shortcut is projected by a unit-width convolution. The body is followed by dropout, global-average pooling over the frequency axis, and a dense layer with a softmax activation. The convolutions use a kernel of $51$ samples throughout, spanning $\approx 1.6\,\mu$Hz at the training resolution, and the dropout rate is $0.1$. The network has $6.0\times 10^{6}$ trainable parameters.

Global average pooling is appropriate here because $\nu_{\max}$ is encoded in the \emph{shape} of the envelope rather than in its position alone: the pooled features summarise the spectrum as a whole and the dense layer maps them to the output bins. This is not the case for the $\Delta\nu$ module, where the position of a feature along the input axis is the quantity being measured, and the head is constructed differently (Sect.~\ref{sec:dnu-architecture}).

\subsubsection{Output binning and training}
\label{sec:binning}

Following \citet{dhanpal2022, dhanpal2023} and \citet{ghanghas2026periodspacingsglobalseismic}, we reformulate regression as a classification over discrete bins. The range $4$--$278\,\mu$Hz is divided into bins of width $1\,\mu$Hz and the final dense layer carries one softmax unit per bin, so that the network returns a normalised probability distribution over $\nu_{\max}$ rather than a single number. Trained under the categorical cross-entropy loss, the outputs of such a classifier approximate the Bayesian posterior probabilities of the bins \citep{richard1991_nn_bayesian_dist}; the resulting distribution is the basis for point estimates, uncertainties and quality diagnostics of Sect.~\ref{sec:uncertainties}. Discretising the output also enforces positivity and admits multimodal solutions, both of which are relevant when the power excess is weak or when a harmonic of the true value competes with it.

Targets are one-hot-encoded bin indices and the loss is categorical cross-entropy. The network is optimised with Adam at an initial learning rate of $1.3\times10^{-3}$ and a batch size of $128$. The learning rate is halved if the validation loss fails to improve for two consecutive epochs, allowing the optimiser to take finer steps once progress slows. Training is stopped after five epochs without improvement to prevent overfitting, and weights are restored to those of the best epoch to obtain the best-performing model. The adopted model converged after $28$ epochs on a single GPU, with the best validation loss reached at epoch $23$.

\subsection{The $\Delta\nu$ module}
\label{sec:dnu-module}

The large separation is a periodicity in frequency, and is therefore most directly accessible in the power spectrum of the power spectrum. The module takes as input a PSPS computed over a window centred on $\nu_{\max}$ returned by the first module, interpolated onto a fixed logarithmic grid in $\Delta\nu$.

\subsubsection{Construction of the PSPS}
\label{sec:psps}

The PSD is first divided by its maximum, then flattened by a background estimated with a sliding median filter in $\log\nu$. The filter uses a window of $0.25$~dex advanced in half-window steps, and the median is converted to a mean by dividing by $\ln 2$, the appropriate correction for a $\chi^2_2$ distribution, so that the flattened spectrum $S(\nu)$ has an expectation value of unity in the absence of signal.

A window of full width $W = 2\,\delta\nu_{\rm env}$ is then extracted around $\nu_{\max}$, where
\begin{equation}
  \delta\nu_{\rm env} = 0.66\,\nu_{\max}^{0.88}\ \mu{\rm Hz}
  \label{eq:fwhm}
\end{equation}
is the envelope full width at half maximum \citep{Mosser_2012_power_excess}. The window is clipped to the observed frequency range. Within it, a broad trend is removed by subtracting a Gaussian-smoothed version of the crop with a standard deviation equal to one tenth of its length, which suppresses the residual envelope shape that would otherwise dominate the low-frequency end of the PSPS. The detrended crop is lightly smoothed with a Gaussian of width $0.1\,\mu$Hz -- specified in physical units rather than in samples, so that its effect is independent of the frequency resolution -- and tapered with a Hann window to suppress edge ringing. The PSPS is the squared modulus of the real discrete Fourier transform of the tapered crop, smoothed by one sample to merge peaks split across adjacent bins.

\begin{figure*}[htbp]
  \centering
  \begin{subfigure}[t]{0.49\textwidth}
    \includegraphics[width=\linewidth]{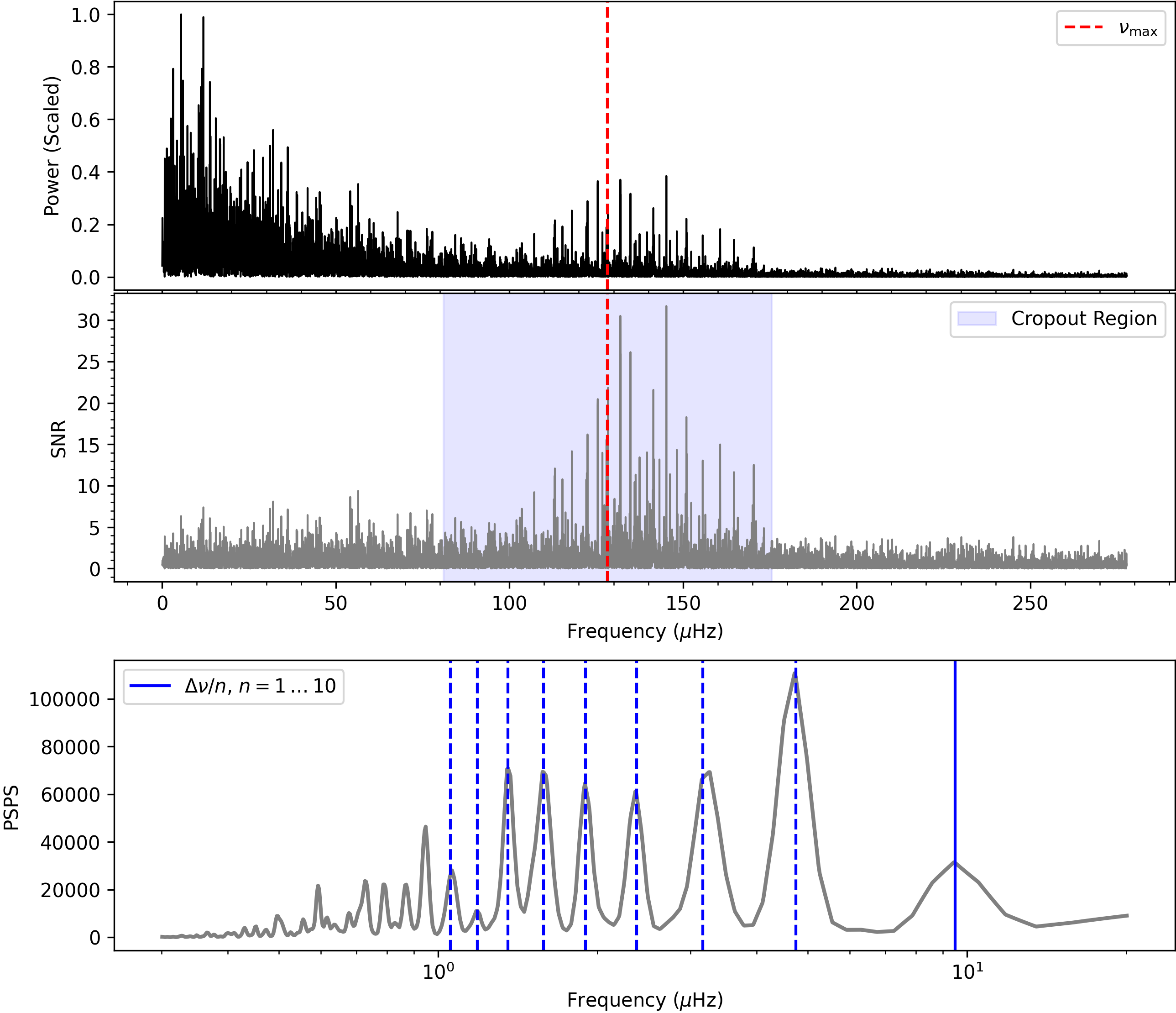}
\caption{High signal-to-noise}
  \end{subfigure}\hfill
  \begin{subfigure}[t]{0.49\textwidth}
    \includegraphics[width=\linewidth]{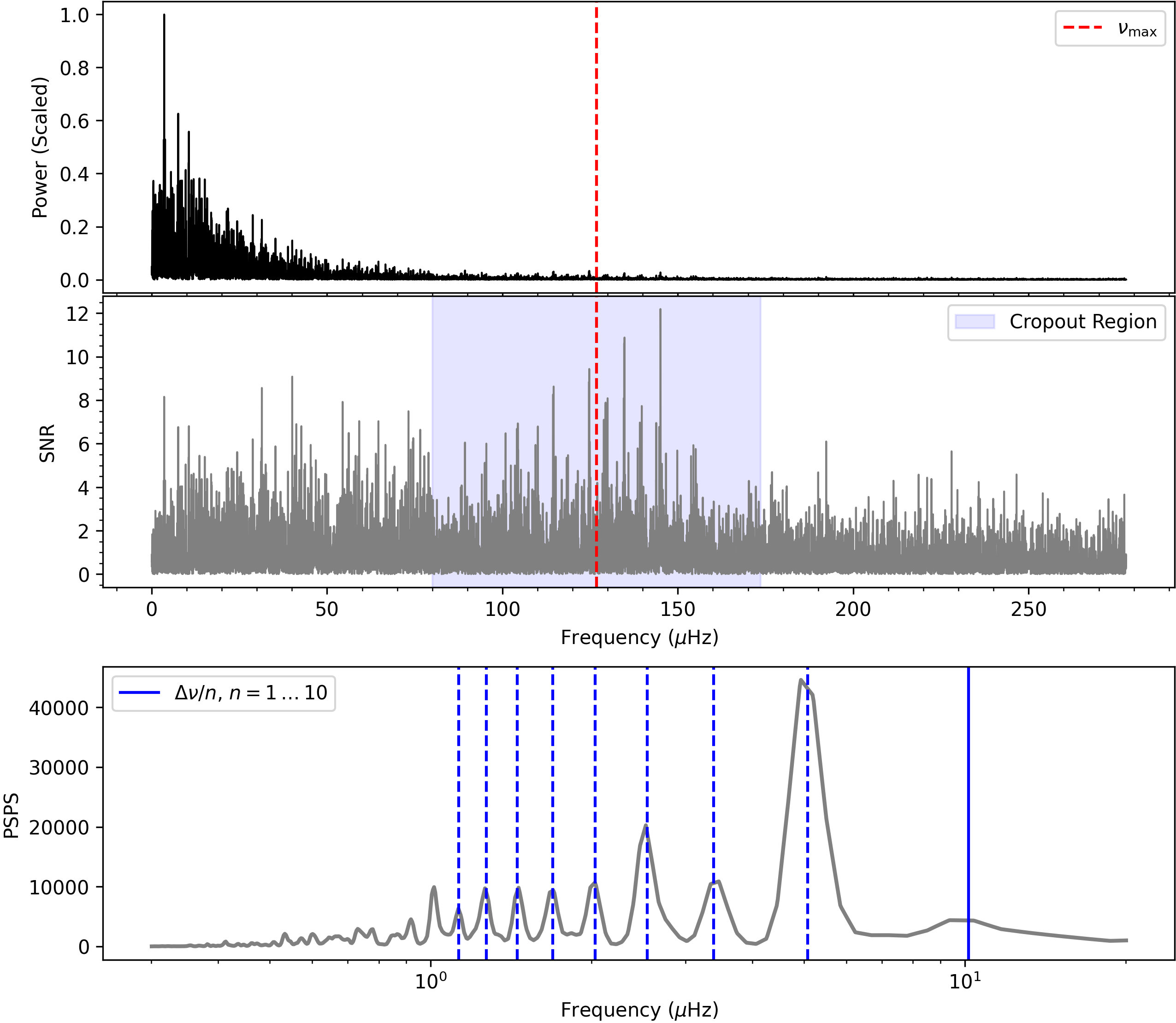}
\caption{Low signal-to-noise}
  \end{subfigure}
\caption{PSPS for two synthetic spectra with closely matched global parameters, $\nu_{\max} \approx 127\,\mu$Hz in both cases and $\Delta\nu = 9.3$ and $10.2\,\mu$Hz respectively, differing in signal-to-noise. \emph{Top:} the power spectrum, normalised by its maximum, with $\nu_{\max}$ marked. \emph{Middle:} the same spectrum after division by the sliding logarithmic median, with the window of width $2\,\delta\nu_{\rm env}$ that is extracted around $\nu_{\max}$ shaded. \emph{Bottom:} the resulting PSPS on the fixed logarithmic grid of Sect.~\ref{sec:dnu-grid}, with $\Delta\nu$ (solid) and the subharmonics $\Delta\nu/n$ for $n = 2 \dots 10$ (dashed) indicated.}
  \label{fig:psps-example}
\end{figure*}

Figure~\ref{fig:psps-example} shows the result for two synthetic spectra chosen to have almost the same $\nu_{\max}$ and $\Delta\nu$ and differ only in signal-to-noise. Note that even in the low signal-to-noise case, the subharmonic structure is still visible, which makes it possible to infer \dnu despite the weak signal. Two further features of the PSPS explain why we infer $\Delta\nu$ with a network rather than by locating a peak.

First, the tallest peak of the PSPS is not at $\Delta\nu$. In both panels it lies at $\Delta\nu/2$. This is a property of the p-mode pattern: the $\ell=0$ and $\ell=2$ modes form closely spaced pairs and the $\ell=1$ modes fall approximately midway between consecutive pairs, so the dominant periodicity of the comb is half the large separation. A method that took the highest peak would return $\Delta\nu/2$ for most red giants. Thus identifying which requires the whole pattern. Furthermore, in the high signal-to-noise case the subharmonic series is clean and the peak at $\Delta\nu$ itself is clearly present, so the correct member can be identified. In the low signal-to-noise case, the peak at $\Delta\nu$ has almost vanished, leaving a strong $\Delta\nu/2$ peak as the first member of the series. 

Second, the harmonics constrain $\Delta\nu$ jointly. Each member of the series carries the same spacing information, so a network reading the whole family does not depend on any single peak being sharp. Where the signal is weak and individual peaks broaden, the series as a whole still fixes $\Delta\nu$. The network is trained on the whole grid for exactly this reason, and the fixed logarithmic axis of Sect.~\ref{sec:dnu-grid} is chosen so that the subharmonic spacing is independent of $\Delta\nu$.

\subsubsection{Fixed logarithmic $\Delta\nu$ grid}
\label{sec:dnu-grid}

The PSPS transform is taken on the crop at its native resolution, so its conjugate axis is a physical-period axis and each sample corresponds to a definite value of $\Delta\nu$. We exploit this by interpolating the PSPS onto a fixed grid in $\Delta\nu$, rather than onto a fixed number of points. The grid is logarithmic with a constant fractional step of $0.5\%$ and contains $832$ points spanning $0.333$--$21.0\,\mu$Hz. It is the input axis of the network, and is stored with the model file.

The grid shares its logarithmic step with the output bins, which cover $1$--$19\,\mu$Hz in $591$ bins of constant fractional width $0.5\%$. The mapping from grid index to class index is therefore a constant shift of $221$ samples rather than an interpolation. The grid extends below the lowest output bin by exactly that shift, which is a factor of three in $\Delta\nu$: this keeps the $\Delta\nu/2$ and $\Delta\nu/3$ sub-harmonics of every admissible solution inside the input, so that the network can see the harmonic structure that distinguishes a genuine comb from an alias. It extends $10\%$ above the highest output bin so that the top class is not located at the edge of the input. On a logarithmic axis, the separation between a peak and its $\Delta\nu/2$ alias is independent of $\Delta\nu$, and equals $139$ grid samples throughout. This helps in allowing the network to use the same convolutional filters to detect this pattern at any $\Delta\nu$.

\subsubsection{Architecture}
\label{sec:dnu-architecture}

The network is a single-branch residual network operating on the $832$-sample PSPS. After batch normalisation and a stem convolution with $32$ filters, the body consists of eight pre-activation residual units in four stages of $[2,2,2,2]$ units with $[32,64,128,256]$ filters. In contrast with the $\nu_{\max}$ module, \emph{all} strides are unity and no pooling is applied along the input axis: because the grid is calibrated in $\Delta\nu$, every downsampling step would halve the resolution in $\Delta\nu$ available to the output layer. The residual units are therefore widened without being shortened, which requires the shortcut to be projected whenever the width changes rather than only when the stride does.

The head is likewise position-aware. The body output is normalised and activated, reduced from $256$ to $8$ channels by a unit-width convolution, and flattened -- retaining the $\Delta\nu$ axis, which a global pooling operation would discard -- before a dense softmax layer over the $591$ output bins. The network has $7.3\times10^{6}$ trainable parameters. The kernel is $11$ samples, which for a $0.5\%$ grid step spans a fractional range of $\approx 5.6\%$ in $\Delta\nu$, and the dropout rate is $0.2$.

\subsubsection{Soft labels and training}

Because the output bins are narrow -- $0.5\%$ in $\Delta\nu$, comparable to the precision the network can be expected to reach -- a one-hot target would place all of the probability mass in one bin and treat an error of one bin as equivalent to an error of a hundred. We therefore train against Gaussian soft labels in $\log\Delta\nu$,
\begin{equation}
  w_k \propto \exp\left[
    -\frac{\left(\log_{10}\Delta\nu - \log_{10} c_k\right)^2}
          {2\,\left(\sigma_b\,\delta_{\log}\right)^2}\right],
  \qquad \sum_k w_k = 1,
  \label{eq:soft-labels}
\end{equation}
where $c_k$ is the centre of bin $k$, $\delta_{\log}$ the constant logarithmic bin step and $\sigma_b = 1$ the label width in units of bins. The Gaussian is constructed in $\log\Delta\nu$ rather than in $\Delta\nu$. On logarithmically spaced bins, a label built in linear units collapses to a one-hot vector at the top of the range. The loss is the categorical cross-entropy against these targets.

The optimiser settings mirror those of the $\nu_{\max}$ module, with an initial learning rate of $1.3\times10^{-3}$, a batch size of $128$, a learning-rate reduction by a factor of ten on a two-epoch plateau, and early stopping with a patience of four epochs and restoration of the best weights.

Alongside the bin-identity accuracy, which is a poor guide on bins this narrow, we monitor two metrics defined in physical units: the fraction of validation spectra whose predicted $\Delta\nu$ lies within a given tolerance of the true value, and the median absolute fractional error. Both are accumulated over the full epoch rather than averaged over batches. The adopted model converged after $10$ epochs on a single GPU, with the best validation loss reached at epoch $6$. At that point the median absolute fractional error was $0.27\%$, and $90\%$ of the validation spectra were predicted within $1\%$ of their true $\Delta\nu$.

\subsubsection{A multi-input alternative}
\label{sec:pspsM-multi}

We also developed and tested a multi-input variant of this module, in which the cropped PSD and its PSPS were supplied as two parallel convolutional branches and the physical width of the window and the scaled $\nu_{\max}$ were supplied as a third, dense branch, the three being concatenated before the classifier. The motivation was to give the network the envelope itself alongside its periodicity, and to tell it explicitly what frequency range the crop covered. In that design both spectral branches were interpolated onto a fixed number of samples rather than onto a physical axis, which makes the mapping from sample index to $\Delta\nu$ depend on the frequency resolution of the input (Sect.~\ref{sec:resolution}). We found no advantage over the single-branch model that would justify this additional coupling, and all results reported here use the fixed-grid single-branch module.

\subsection{Matching the frequency resolution at inference}
\label{sec:resolution}

Both modules are trained at a single frequency resolution, $\delta\nu_{\rm tr} = 0.0317\,\mu$Hz, and both are sensitive to a departure from it, although for different reasons and with very different consequences.

For the $\nu_{\max}$ module the input is a fixed frequency grid, so a spectrum at a different resolution is simply resampled onto it. The resampling is harmless for our purpose: at higher resolution the power is averaged, which reduces the variance of the $\chi^2_2$ noise in a way the network has not seen, but does not displace any feature.

For a PSPS the situation is more delicate, and in the multi-input design of Sect.~\ref{sec:pspsM-multi} it can produce a large error. Taking the transform on the native-resolution crop gives a conjugate axis spanning $[0,\,1/(2\delta\nu)]$ in period, so the physical meaning of a given sample is proportional to $\delta\nu$. Interpolating that array onto a fixed \emph{number} of samples normalises its length but not its span, with the result that a spectrum supplied at a different resolution places its $\Delta\nu$ peak at a different sample and the inferred $\Delta\nu$ is wrong by the ratio of the two resolutions. A four-year \emph{Kepler} PSD at $\delta\nu = 0.0079\,\mu$Hz analysed by a module trained at $0.0317\,\mu$Hz yields a $\Delta\nu$ too small by a factor of four. Nothing downstream detects this: the input shapes are identical, the stored window width is identical, and the output distribution is narrow and confident.

The fixed logarithmic grid of Sect.~\ref{sec:dnu-grid} removes the failure by construction, because the interpolation is performed against the crop's own period axis and the input axis is therefore physical. A resolution mismatch then degrades the PSPS without rescaling it: the smoothing widths of Sect.~\ref{sec:psps} are set in physical units and in fractions of the crop length, so they are reproduced correctly, but the noise statistics of the crop are not those the network was trained on. The module nevertheless verifies the resolution of every spectrum it is given and resamples any that departs from $\delta\nu_{\rm tr}$ by more than $1\%$ onto the training grid before constructing the PSPS.

Two safeguards complete the inference path. The first is the resolution check just described. The second rejects, rather than predicts on, spectra that cannot be processed meaningfully: those containing non-finite power, those for which the $\nu_{\max}$ module returned an unusable value, those whose window around $\nu_{\max}$ contains fewer than $32$ native samples -- below which the detrending of Sect.~\ref{sec:psps} degenerates and the PSPS carries no information on $\Delta\nu$ -- and those whose window is featureless, so that the detrended crop is identically zero and the normalisation is undefined.

Neither safeguard is exercised by the data analysed in this work. The \emph{TESS} power spectra are computed on the one-year grid directly, and the \emph{Kepler} spectra are degraded to that baseline before analysis (Sect.~\ref{sec:data}), so every spectrum presented to the $\Delta\nu$ module is already at $\delta\nu_{\rm tr}$ and none was resampled; no spectrum was rejected in any of the runs reported here. We describe both because they are required for the module to be applied to spectra at a different resolution -- a shorter TESS baseline, or the full four-year \emph{Kepler} series -- which is the use the modular design is intended to support.

\subsection{Point estimates, uncertainties and reliability}
\label{sec:uncertainties}

Each module returns a vector of probabilities $p_k$ over its $N_b$ output bins. Bin edges are reconstructed from the centres by taking the midpoints of adjacent centres internally and extrapolating symmetrically at the two ends. From the cumulative distribution we read the $16$th, $50$th and $84$th percentiles by linear interpolation against the edges, and adopt the median $x_{50}$ as the point estimate with asymmetric uncertainties $\sigma_- = x_{50}-x_{16}$ and $\sigma_+ = x_{84}-x_{50}$. We write $\bar\sigma = (\sigma_-+\sigma_+)/2$ for the mean half-width, and quote fractional uncertainties as $\bar\sigma/x_{50}$. We also record the mode, the centre of the most probable bin.

Two scalar summaries of the shape of the distribution are computed. The confidence $C = \max_k p_k$ is the probability of the most probable bin. The normalised entropy
\begin{equation}
  \mathcal{H} = -\frac{1}{\ln N_b}\sum_k p_k \ln p_k
  \label{eq:entropy}
\end{equation}
is zero for a distribution concentrated in one bin and unity for a uniform one; it measures how much of the range the distribution spans, independently of where the probability sits.

Neither quantity alone is sufficient. A distribution may be sharply peaked yet carry a large uncertainty if the peak is broad relative to the bin width, and a narrow interval means little if the bins are wide. We therefore combine them into a reliability index,
\begin{equation}
  \mathcal{R} = C\left(1 - \frac{\bar\sigma}{T}\right),
  \qquad
  T = \sqrt{\left(\frac{w}{2}\right)^{2} + \left(f\,x_{50}\right)^{2}},
  \label{eq:ri}
\end{equation}
where $w$ is the median bin width and $f = 0.1$. The threshold $T$ is the uncertainty that a prediction of this magnitude could attain if it were limited either by the discretisation, through $w/2$, or by a target fractional precision of $f$, whichever dominates. A prediction whose uncertainty is well below $T$ and whose probability is concentrated has $\mathcal{R}$ approaching unity; a prediction whose uncertainty reaches $T$ has $\mathcal{R} \le 0$ regardless of how confident the network is.

Three further conditions flag distributions that the summary statistics above describe poorly. A distribution with more than $16\%$ of its mass in the lowest or highest bin is \emph{truncated}: the network is indicating a value at or beyond the edge of the trained range, and the percentiles are then bounded by that range rather than by the data. A distribution whose mode falls outside the $16$th--$84$th percentile interval is \emph{inconsistent}: this arises when the output is strongly bimodal and the median falls in the low-probability trough between two competing peaks, so that it represents neither. A distribution with $\mathcal{H} > 0.9$ is \emph{flat}. We refer to a prediction satisfying either of the first two conditions as out of bounds, and exclude such predictions in Sect.~\ref{sec:selection}.

\subsection{Spurious-signal vetting}
\label{sec:spurious}

The modules return a prediction for every spectrum, including spectra whose dominant feature is not a solar-like oscillation envelope at all. Classical pulsators, contaminating binaries and instrumental artefacts all produce power excesses that a network trained only on red giants cannot recognise as foreign, and which can yield narrow, confident and entirely spurious values of $\nu_{\max}$. We therefore apply an independent morphological test to each spectrum, derived from the peak structure of the background-divided spectrum rather than from the network output.

The test examines the ten tallest peaks of the signal-to-noise spectrum and tests whether their heights, spacings and concentration are compatible with a stochastically excited Gaussian envelope. A spectrum whose tallest peak is too high to be stochastic, whose second peak falls far below the first, whose comparable peaks are spread more widely than an envelope of the appropriate width could contain, or whose peaks follow a multiplicative rather than an additive comb, is flagged as spurious. A spectrum whose tallest peak is consistent with $\chi^2_2$ noise is flagged as indeterminate, carrying no information either way, and the remainder are flagged as clean. The individual criteria and their thresholds are collected in Appendix~\ref{app:spurious}.

Two caveats apply. The thresholds are empirical: they are set a priori from the noise statistics and from the dynamic range of the flattened spectra, and are not calibrated against a labelled sample, so the test has no known completeness or contamination rate and should not be read as a classifier with quantified performance. They are also tuned for \emph{TESS} spectra at the one-year resolution of Sect.~\ref{sec:data-tess}, and we apply the vetting to that sample alone. The synthetic spectra contain no contaminants by construction, and the \emph{Kepler} sample is a curated list of known red giants, so neither requires it. A trained classifier would be more robust than a hand-tuned decision list and would transfer between instruments and baselines without retuning; we plan to build one.

\subsection{Selection of the reliable sample}
\label{sec:selection}

The quantities defined above are combined into a single selection. For $\nu_{\max}$ we retain predictions satisfying
\begin{enumerate}
  \item $\bar\sigma/\nu_{\max} < 10\%$;
  \item $\mathcal{R} \ge 0.1$;
  \item $\mathcal{H} < 0.7$;
  \item not out of bounds in the sense of Sect.~\ref{sec:uncertainties};
  \item $\nu_{\max} < 260\,\mu$Hz, which keeps the prediction clear of the upper edge of the trained range;
  \item agreement between the two input representations of Sect.~\ref{sec:vmax-module}, $\left|\nu_{\max}^{\rm snr} - \nu_{\max}^{\rm lin}\right| / \nu_{\max}^{\rm lin} < 0.30$;
  \item not flagged as spurious by Sect.~\ref{sec:spurious}.
\end{enumerate}
For $\Delta\nu$ we require all of the above, since the $\Delta\nu$ module is conditioned on $\nu_{\max}$ and inherits its failures, and in addition $\bar\sigma/\Delta\nu < 5\%$, $\mathcal{R} \ge 0.1$, $\mathcal{H} < 0.7$ and not out of bounds, evaluated on the $\Delta\nu$ distribution. The two samples are therefore nested but not identical: a star whose $\nu_{\max}$ is sound but whose $\Delta\nu$ is not is retained in the first and dropped from the second. We quote the two yields separately throughout, since a catalogue used for $\nu_{\max}$ alone should not be limited by the completeness of a parameter it does not use.

Criterion (vi) is the consistency check motivated in Sect.~\ref{sec:vmax-module}. The tolerance of $30\%$ is set so as to sit above the typical differences between the two representations, including the low-frequency shift of the linear module, and well below the factor-level disagreement that a catastrophic failure of the signal-to-noise module produces. It is a veto rather than a correction: a star failing it is removed, not reassigned. We quantify its effect per dataset in Sect.~\ref{sec:res-tess}.

One criterion was deliberately excluded. An earlier version of these criteria required the measured $\nu_{\max}$ to agree with the value implied by the measured $\Delta\nu$ through a scaling relation. That cut is effective, but it imposes a relation on a sample the framework otherwise derives without reference to one, and it biases the $\nu_{\max}$--$\Delta\nu$ plane in which we present the results. With the consistency check of criterion (vi) in place it proved redundant, and we have removed it. No scaling relation enters the selection at any point.

\subsection{Validation metrics}
\label{sec:metrics}

We validate the framework against held-out synthetic spectra, against \emph{Kepler} spectra degraded to the one-year baseline of the training set, and against reference catalogues for the observed samples. For a predicted value $\hat{y}$ and a reference value $y$ we work throughout with the fractional residual $r = (\hat{y}-y)/y$, and summarise its distribution by the median and by the robust dispersion $\sigma_{\rm MAD} = 1.4826\,\mathrm{med}\left|r - \mathrm{med}\,r\right|$, which is insensitive to the tails. We quote separately the fraction of the sample lying beyond $5\,\sigma_{\rm MAD}$ of the median, which measures the catastrophic-failure rate independently of the core of the distribution.

To test whether the reported uncertainties are themselves meaningful we use the normalised residual, formed with the uncertainty on the side facing the reference value,
\begin{equation}
  R =
  \begin{cases}
    \dfrac{\hat{y}-y}{\sqrt{\sigma_+^2 + \sigma_{\rm ref}^2}}, & \hat{y} < y,\\[2ex]
    \dfrac{\hat{y}-y}{\sqrt{\sigma_-^2 + \sigma_{\rm ref}^2}}, & \hat{y} \ge y,
  \end{cases}
  \label{eq:norm-res}
\end{equation}
where $\sigma_{\rm ref}$ is the uncertainty on the reference value. If the uncertainties are well calibrated the fraction of stars with $|R| < 1$ should equal the $68.3\%$ expected for Gaussian errors; a larger fraction indicates under-confidence, a smaller one over-confidence.

\section{Results}
\label{sec:results}

We validate the two modules against three samples. They are presented in order of decreasing control over the reference value, which affects how the numbers should be read. The held-out synthetic spectra are compared against the values used to generate them, so the residual is the error of the module alone. The \emph{Kepler} spectra are compared against \citet{Yu_2018}, measured from the full four-year light curves, while the spectra supplied to the modules are the same stars degraded to the one-year resolution of the training set; the reference is therefore derived from a far longer baseline than the input, and the residual remains dominated by the framework. The \emph{TESS} spectra are compared against an independent analysis of the same stars \citep{Sreenivas_2026}, built from the full continuous-viewing-zone baseline, which is about twice our own. The reference is therefore again the more precise of the two, but by a smaller margin than on \emph{Kepler}, so a larger share of the observed scatter belongs to the reference pipeline.

\begin{figure*}[htbp]
  \centering
  \begin{subfigure}[t]{0.32\textwidth}
    \includegraphics[width=\linewidth]{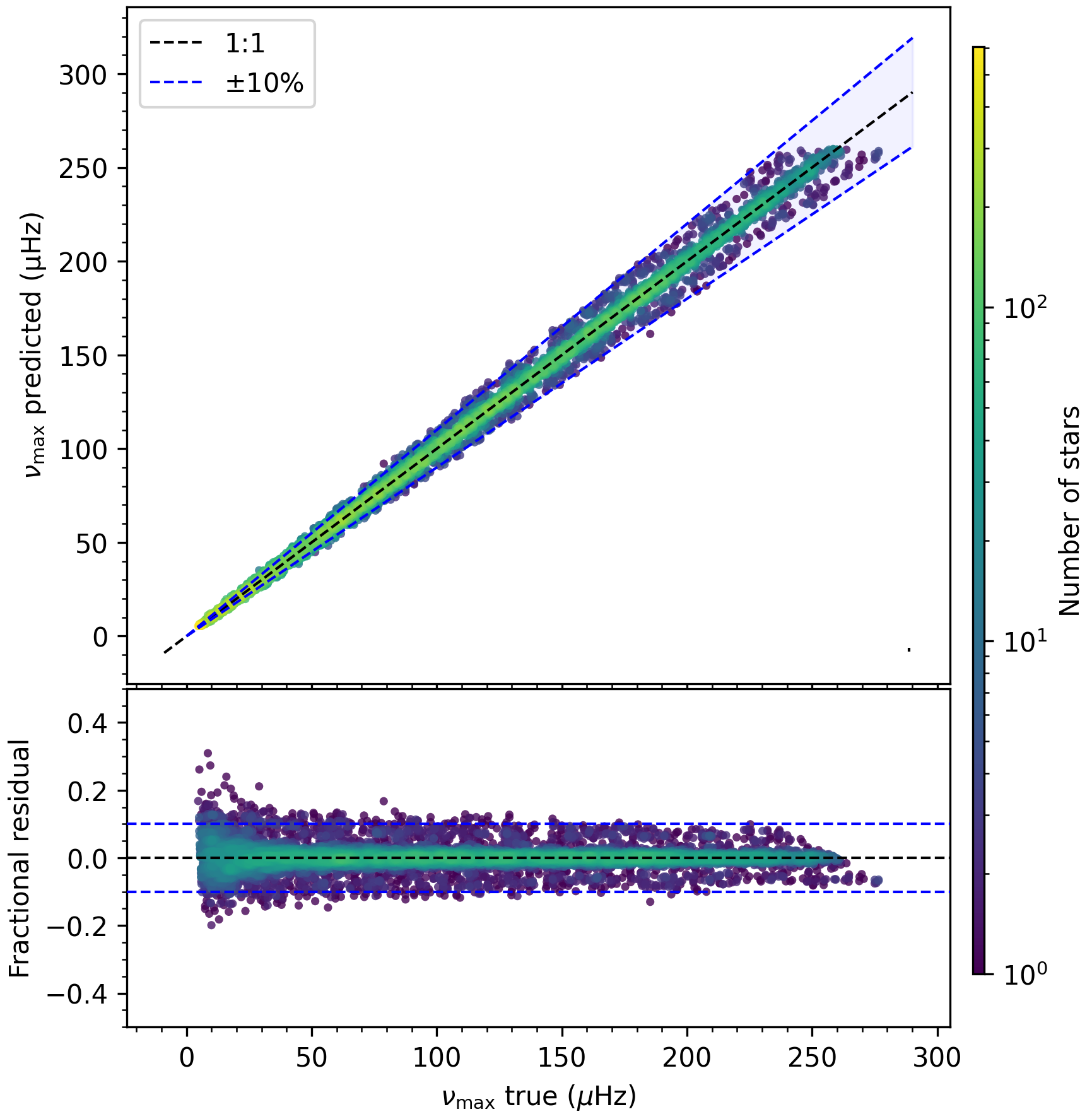}
\caption{Held-out synthetics}
  \end{subfigure}\hfill
  \begin{subfigure}[t]{0.32\textwidth}
    \includegraphics[width=\linewidth]{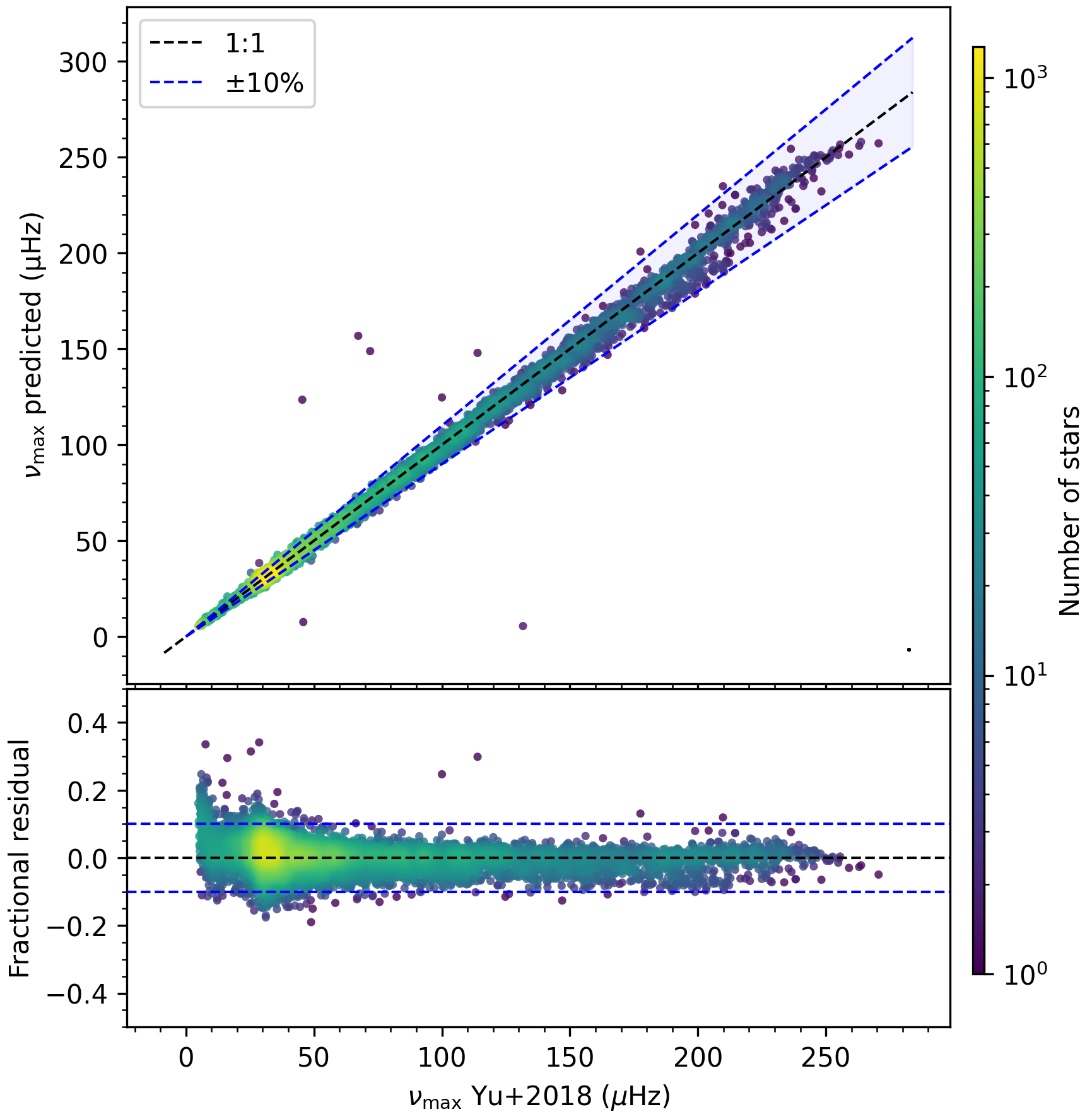}
\caption{\emph{Kepler} at one-year resolution}
  \end{subfigure}\hfill
  \begin{subfigure}[t]{0.32\textwidth}
    \includegraphics[width=\linewidth]{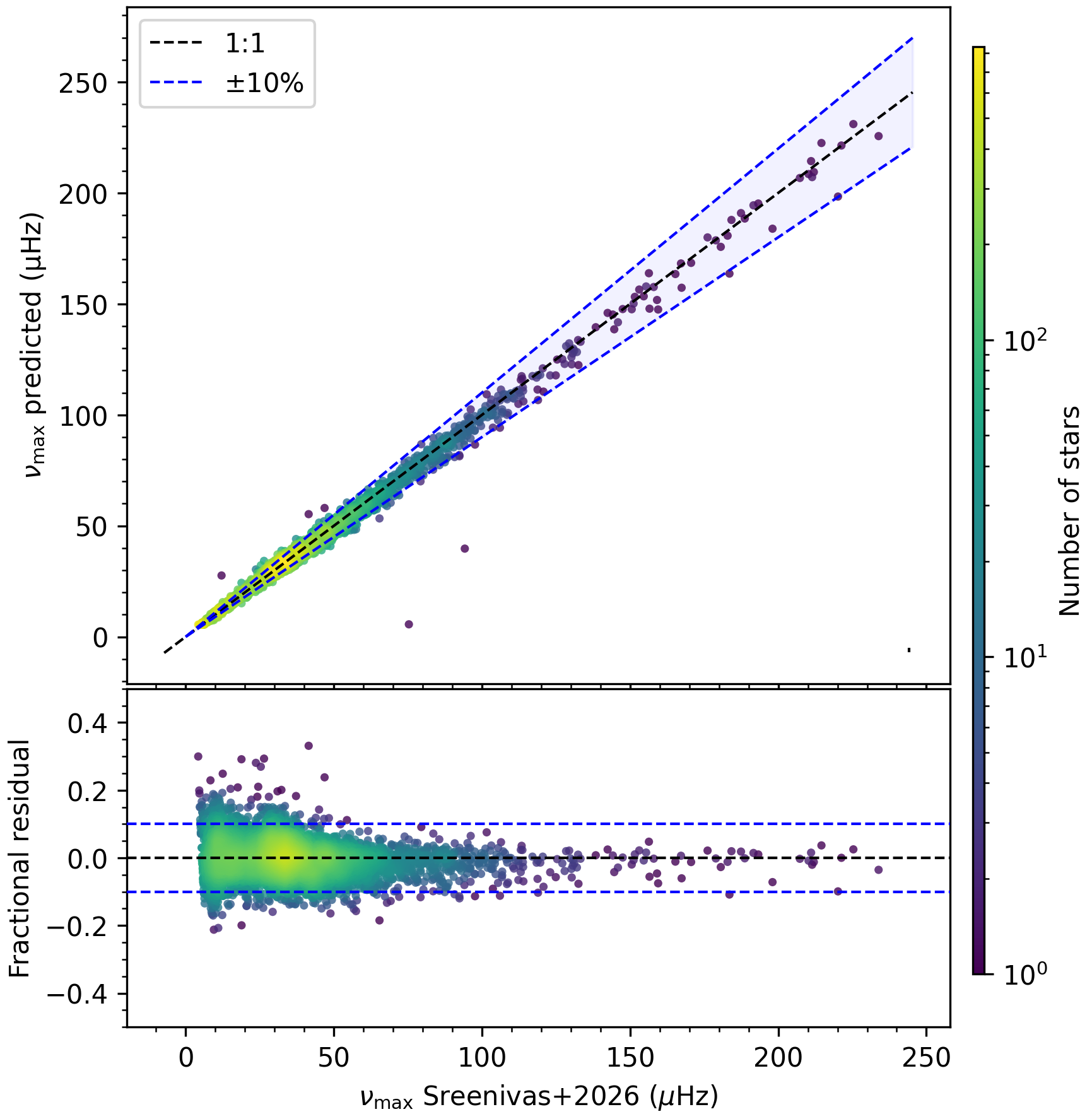}
\caption{\emph{TESS}, reliable reference}
  \end{subfigure}

\vspace{0.4em}

  \begin{subfigure}[t]{0.32\textwidth}
    \includegraphics[width=\linewidth]{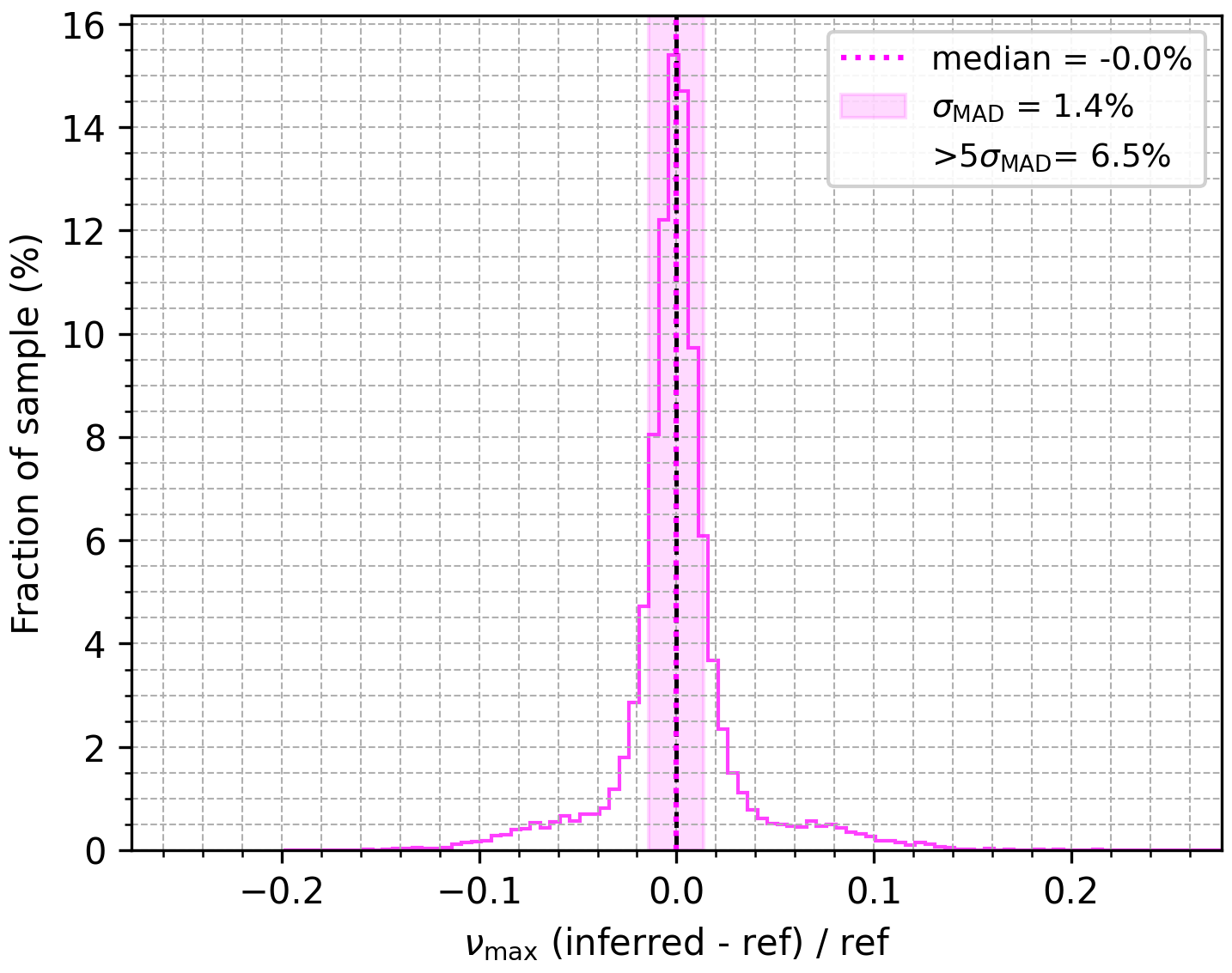}
\caption{}
  \end{subfigure}\hfill
  \begin{subfigure}[t]{0.32\textwidth}
    \includegraphics[width=\linewidth]{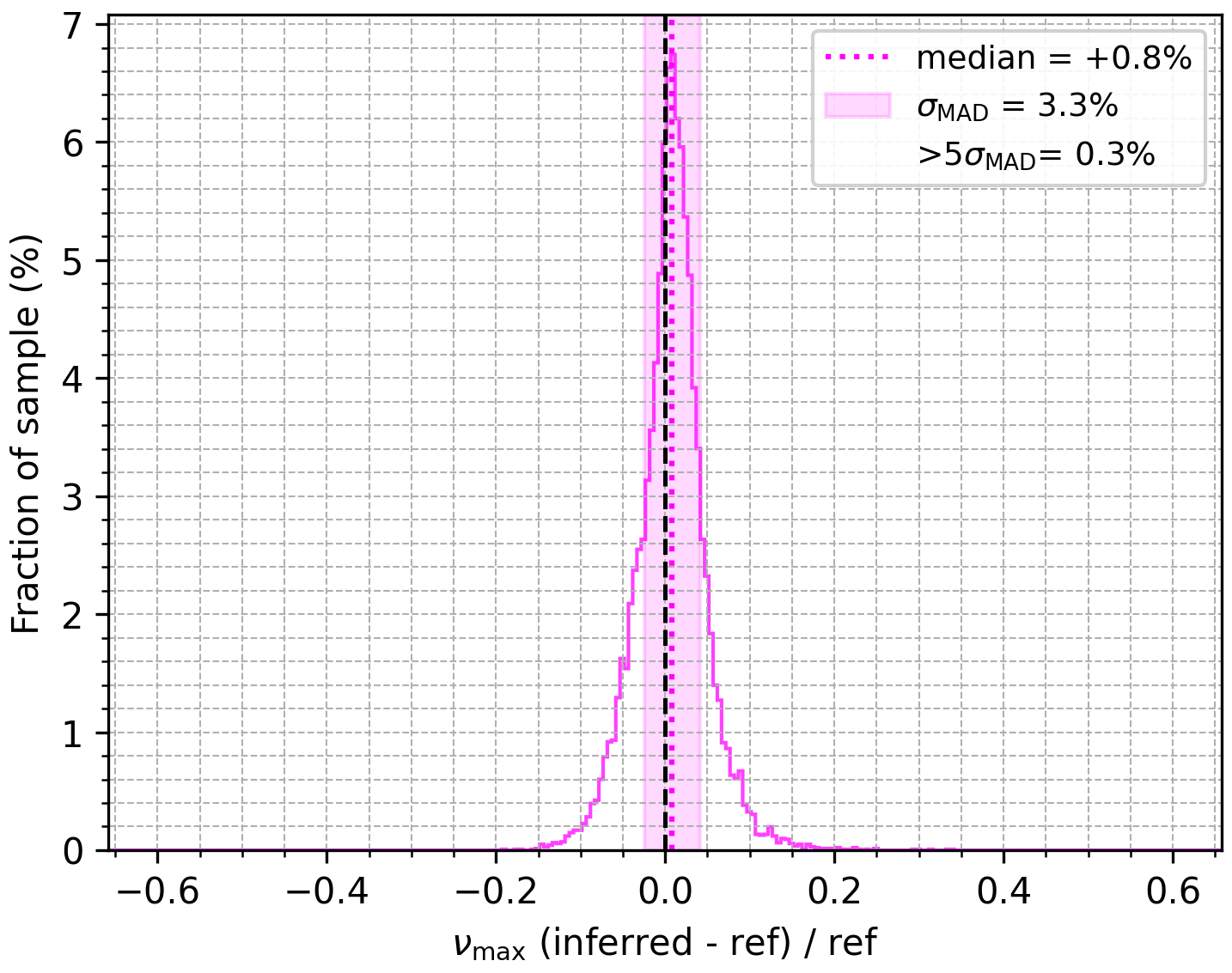}
\caption{}
  \end{subfigure}\hfill
  \begin{subfigure}[t]{0.32\textwidth}
    \includegraphics[width=\linewidth]{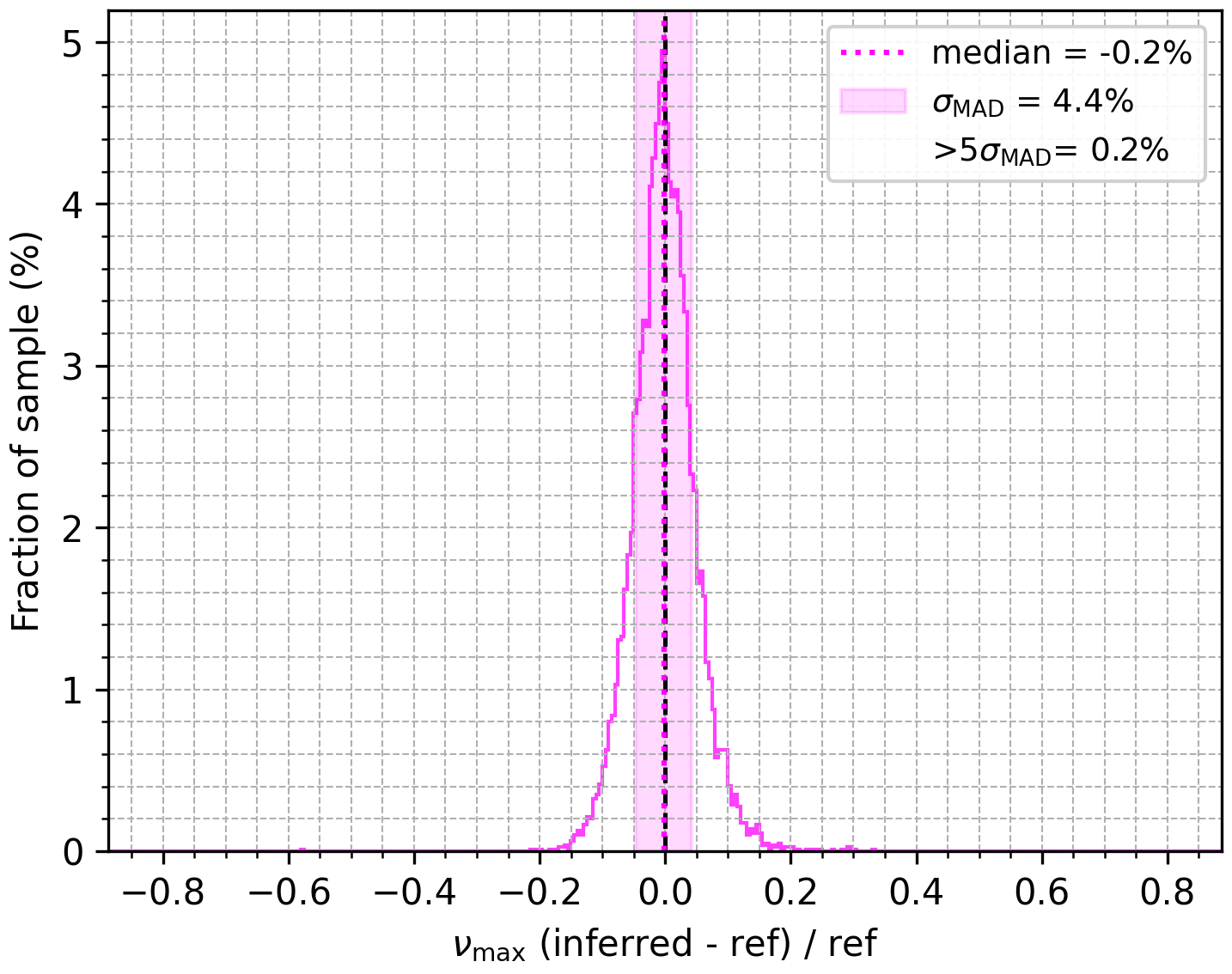}
\caption{}
  \end{subfigure}
\caption{Inferred $\nu_{\max}$ against the reference value for the three validation samples. \emph{Top row:} inferred against reference, coloured by the local number of stars, with the fractional residual below each panel. The black dashed line is the one-to-one relation and the blue dashed lines mark agreement at the $10\%$ level. \emph{Bottom row:} distribution of the fractional residual $r=(\hat{y}-y)/y$ for the same sample, with the median and the robust dispersion $\sigma_{\rm MAD}$ indicated. Note that the horizontal scale differs between the three lower panels. The \emph{TESS} panels are restricted to stars whose reference $\Delta\nu$ is flagged reliable by \citet{Sreenivas_2026}, as discussed in Sect.~\ref{sec:res-tess}.}
  \label{fig:numax-comparison}
\end{figure*}

\begin{figure*}[htbp]
  \centering
  \begin{subfigure}[t]{0.32\textwidth}
    \includegraphics[width=\linewidth]{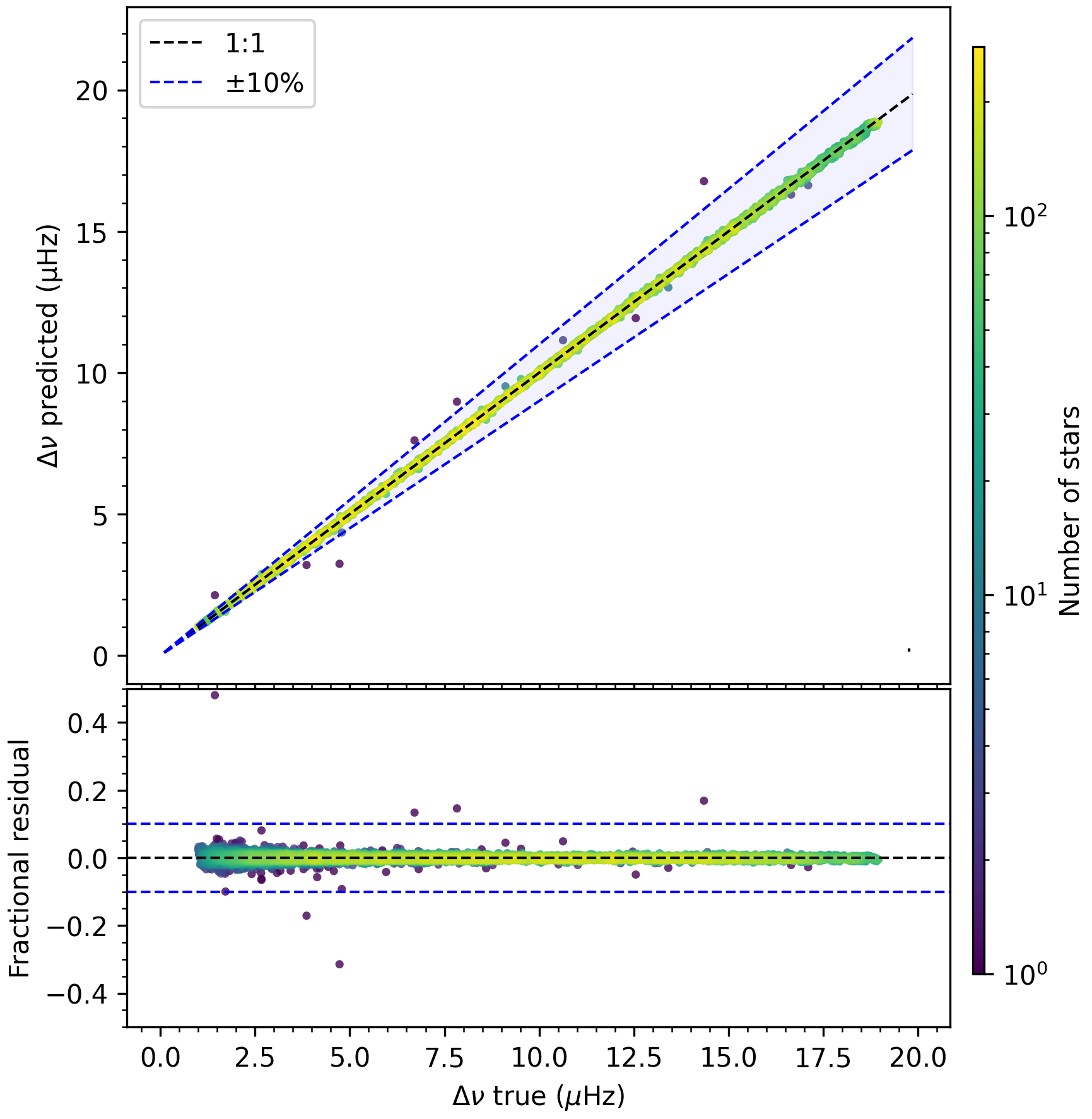}
\caption{Held-out synthetics}
  \end{subfigure}\hfill
  \begin{subfigure}[t]{0.32\textwidth}
    \includegraphics[width=\linewidth]{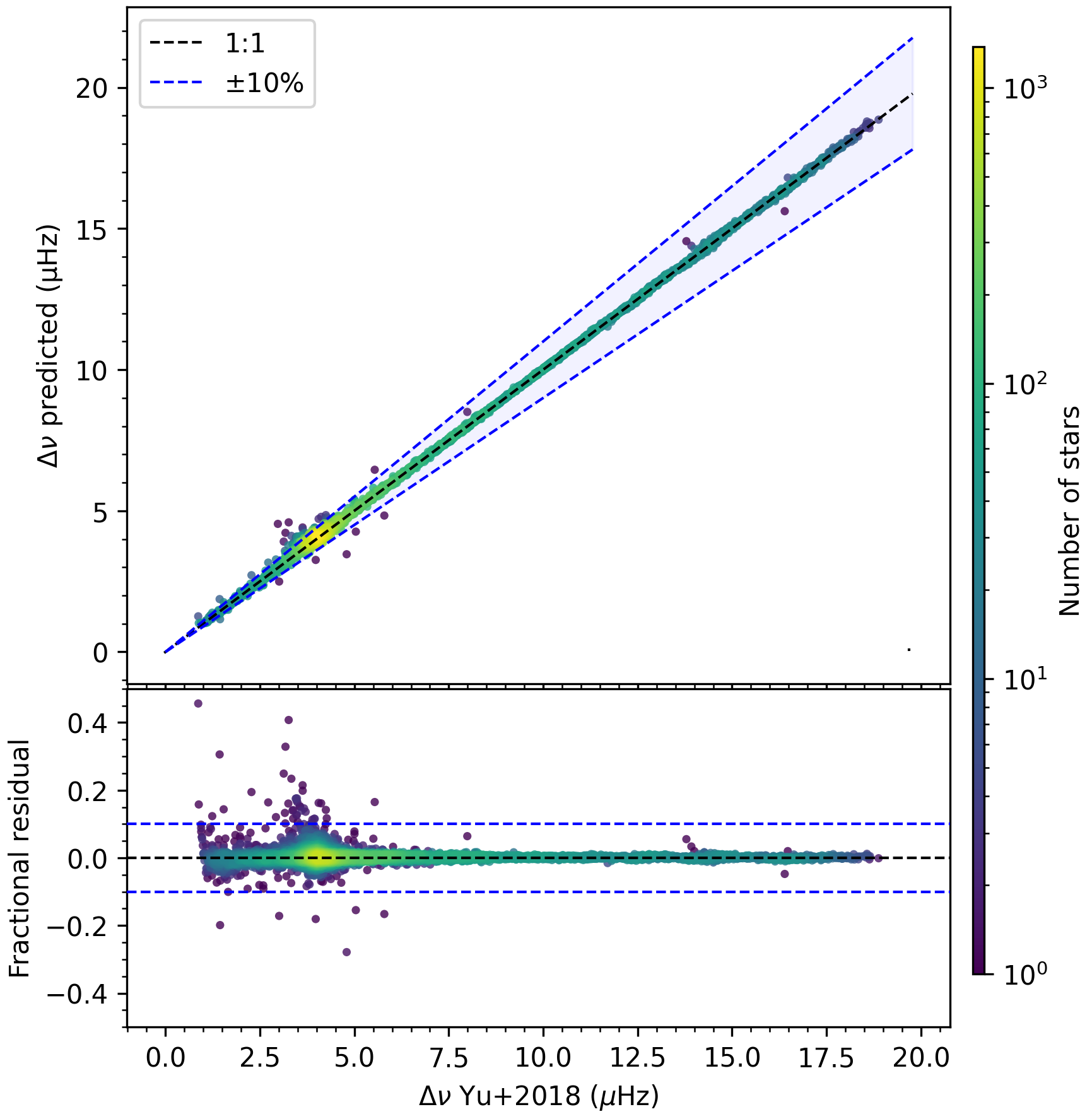}
\caption{\emph{Kepler} at one-year resolution}
  \end{subfigure}\hfill
  \begin{subfigure}[t]{0.32\textwidth}
    \includegraphics[width=\linewidth]{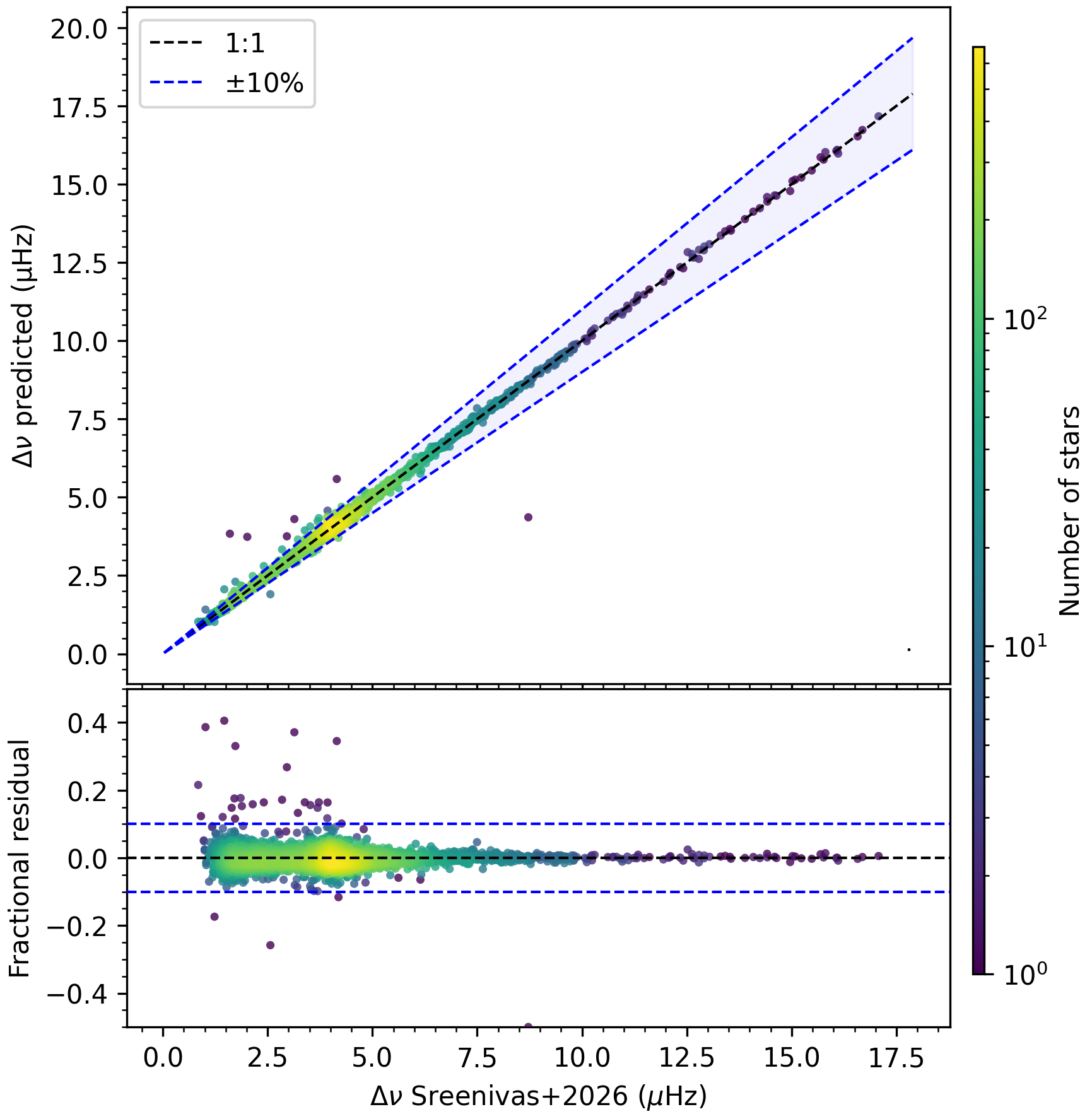}
\caption{\emph{TESS}, reliable reference}
  \end{subfigure}

\vspace{0.4em}

  \begin{subfigure}[t]{0.32\textwidth}
    \includegraphics[width=\linewidth]{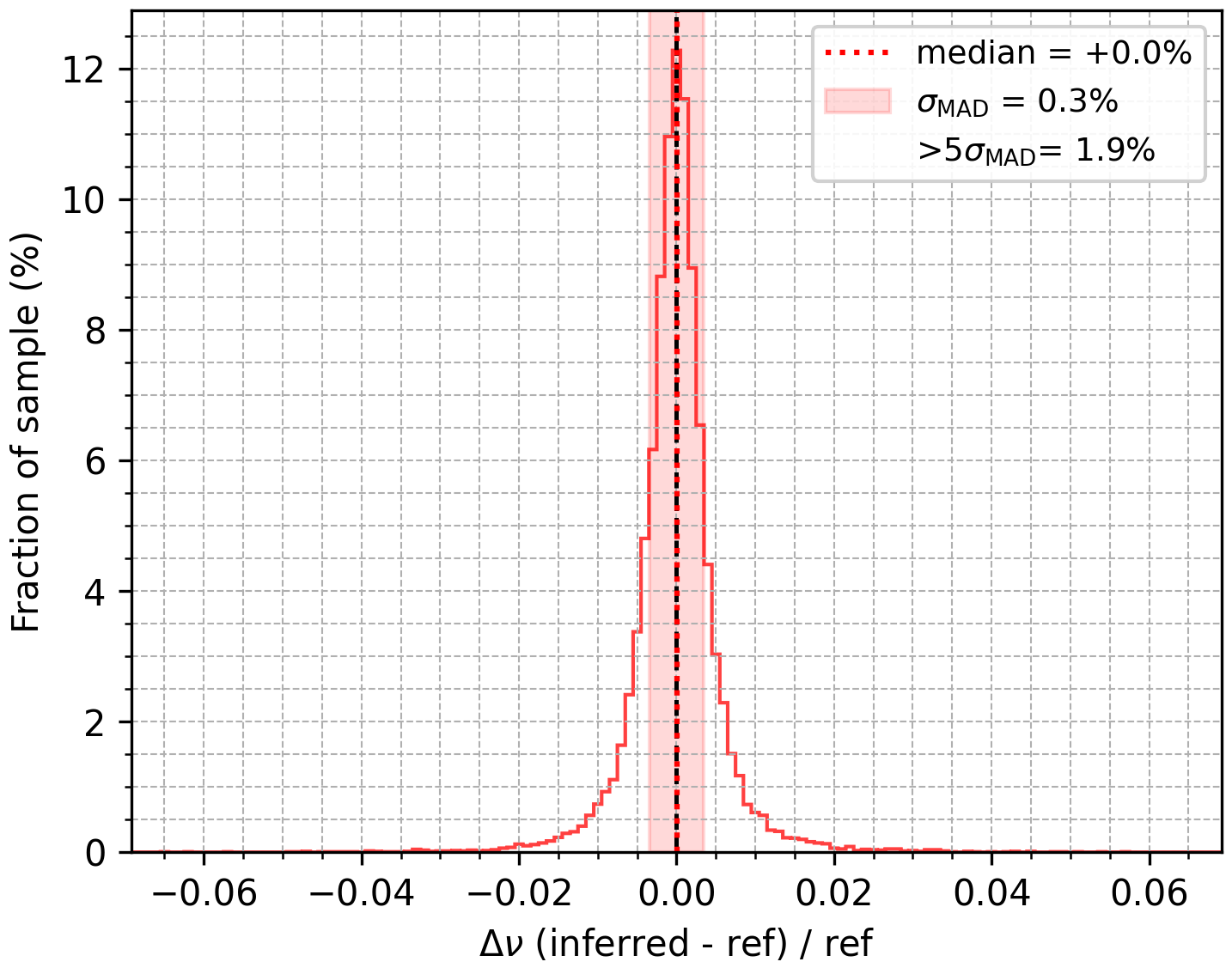}
\caption{}
  \end{subfigure}\hfill
  \begin{subfigure}[t]{0.32\textwidth}
    \includegraphics[width=\linewidth]{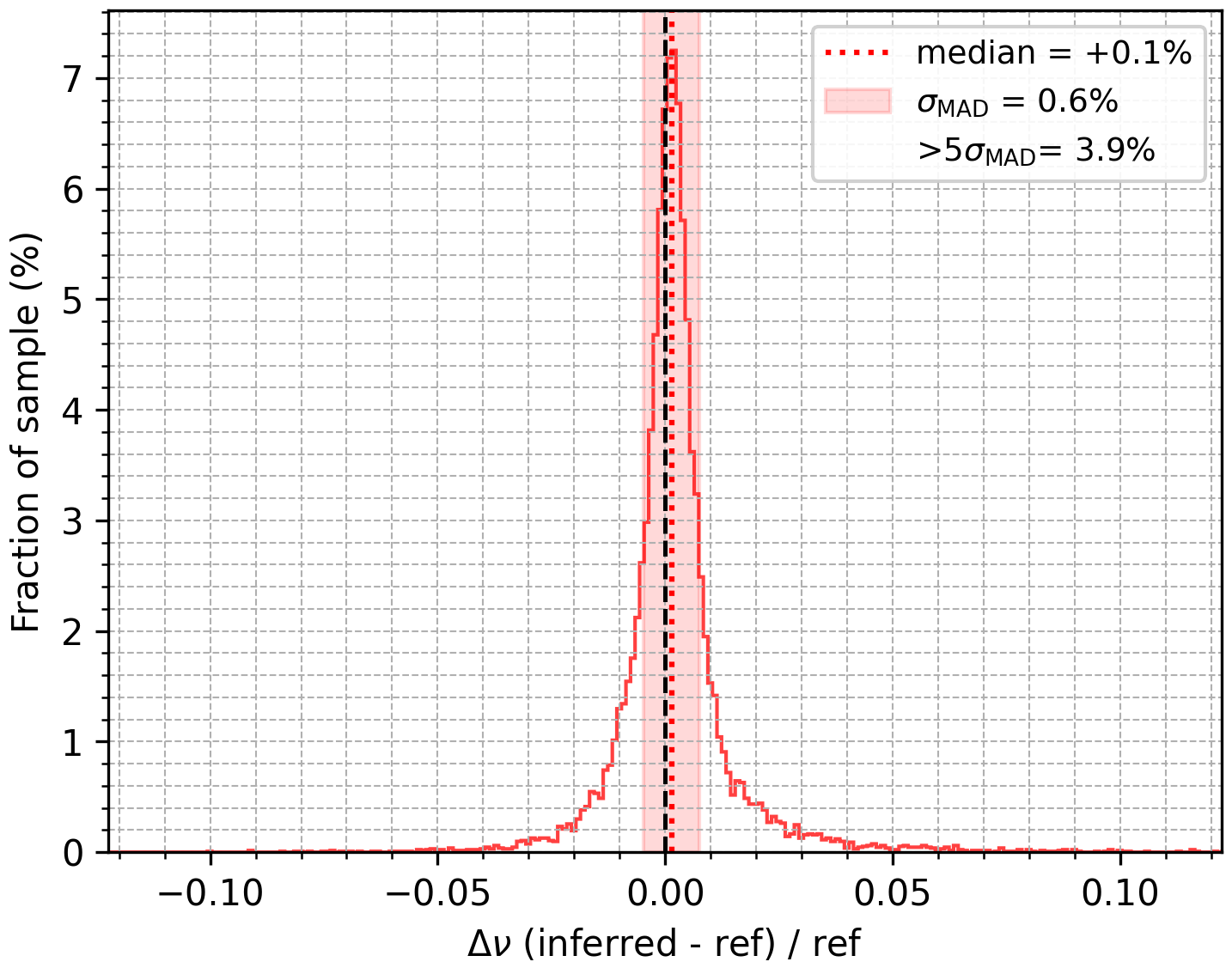}
\caption{}
  \end{subfigure}\hfill
  \begin{subfigure}[t]{0.32\textwidth}
    \includegraphics[width=\linewidth]{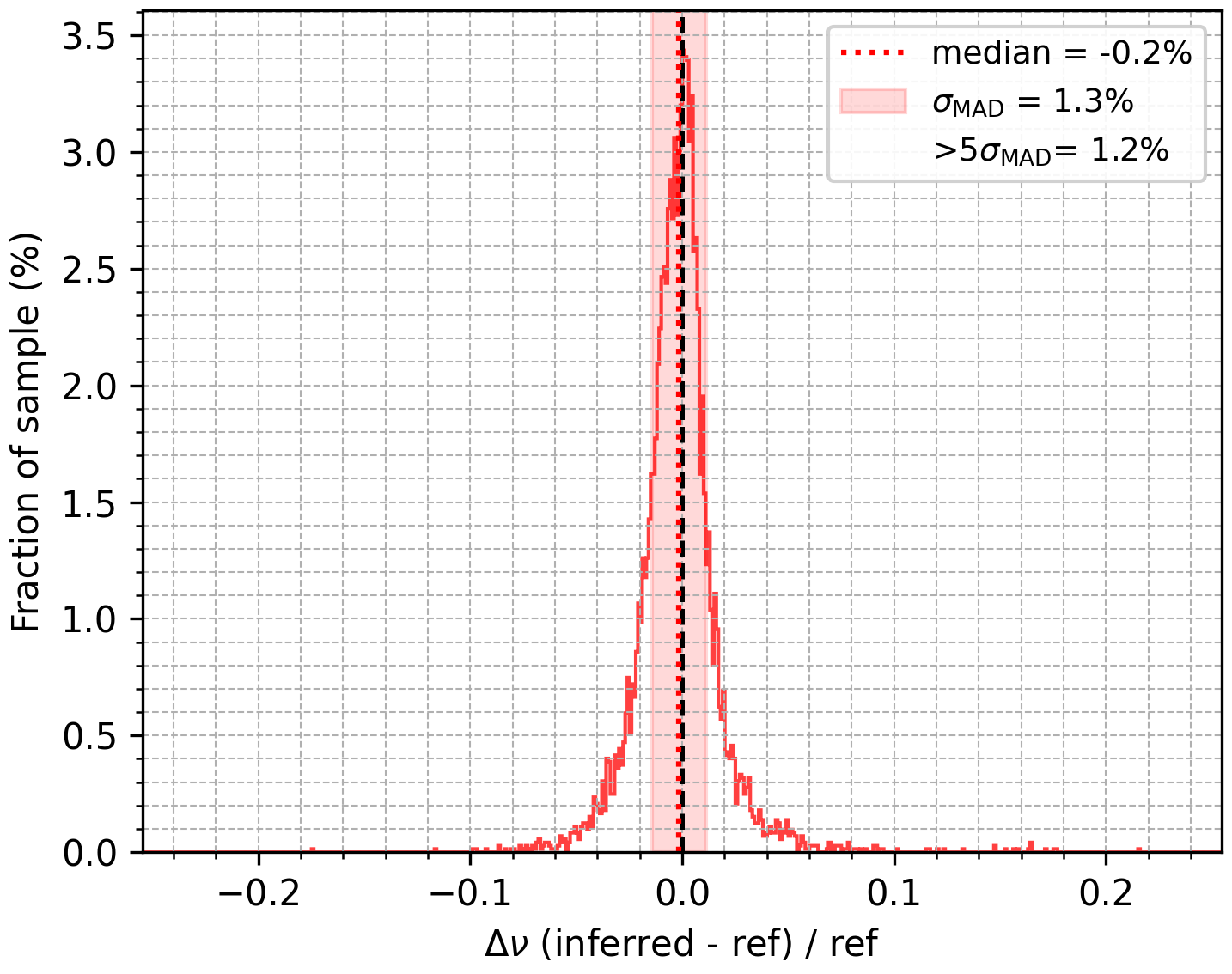}
\caption{}
  \end{subfigure}
\caption{As Fig.~\ref{fig:numax-comparison}, for $\Delta\nu$.}
  \label{fig:dnu-comparison}
\end{figure*}

Figures~\ref{fig:numax-comparison} and~\ref{fig:dnu-comparison} show the three comparisons for $\nu_{\max}$ and for $\Delta\nu$, and Table~\ref{tab:validation} collects the summary statistics. Throughout we use a single reliable sample per dataset for each parameter, defined by the criteria of Sect.~\ref{sec:selection}, so that the $\nu_{\max}$ and $\Delta\nu$ columns of Table~\ref{tab:validation} refer to nested but not identical samples. For $\nu_{\max}$ the selection retains $86\%$ of the synthetic spectra, $90\%$ of the \emph{Kepler} spectra and $79\%$ of the \emph{TESS} spectra; for $\Delta\nu$, which additionally requires its own distribution to be well behaved, the figures are $83\%$, $84\%$ and $72\%$. That \emph{TESS} is the lowest in both is the first indication that it is the harder case.

% ---------------------------------------------------------------- numax figure

% ------------------------------------------------------------------ dnu figure

% ------------------------------------------------------------------- table
\begin{table*}[htbp]
\caption{Summary of the three validation comparisons.}
  \label{tab:validation}
  \centering
  \begin{tabular}{llcccccc}
    \toprule
    Sample & Reference & $N$ & med.\ $r$ & $\sigma_{\rm MAD}$ & $>5\sigma_{\rm MAD}$ & $f_{10\%}$ & $|R|<1$ \\
    \midrule
    \multicolumn{8}{l}{$\nu_{\max}$}\\
    Synthetics    & true values            & $16\,851/19\,523$ & $-0.01\%$ & $1.38\%$ & $6.5\%$  & $98.1\%$ & $65.4\%$ \\
    \emph{Kepler} & \citet{Yu_2018}        & $14\,164/15\,799$ & $+0.81\%$ & $3.29\%$ & $0.28\%$ & $97.3\%$ & $63.0\%$ \\
    \emph{TESS}   & \citet{Sreenivas_2026} & $7981/10\,073$    & $-0.22\%$ & $4.43\%$ & $0.15\%$ & $95.0\%$ & $66.6\%$ \\
    \quad all cross-matched & \citet{Sreenivas_2026} & $12\,427/17\,317$ & $-0.08\%$ & $4.86\%$ & $0.89\%$ & $92.4\%$ & $65.3\%$ \\
    \midrule
    \multicolumn{8}{l}{$\Delta\nu$}\\
    Synthetics    & true values            & $16\,295/19\,519$ & $+0.00\%$ & $0.35\%$ & $1.9\%$  & $100.0\%$ & $87.8\%$ \\
    \emph{Kepler} & \citet{Yu_2018}        & $13\,339/15\,794$ & $+0.14\%$ & $0.61\%$ & $3.9\%$  & $99.6\%$  & $85.6\%$ \\
    \emph{TESS}   & \citet{Sreenivas_2026} & $7220/10\,070$    & $-0.16\%$ & $1.27\%$ & $1.2\%$  & $99.6\%$  & $65.2\%$ \\
    \quad all cross-matched & \citet{Sreenivas_2026} & $10\,423/17\,198$ & $-0.01\%$ & $1.84\%$ & $11.0\%$ & $89.6\%$  & $55.3\%$ \\
    \bottomrule
  \end{tabular}
\tablefoot{$N$ gives the size of the reliable sample and the number of stars with a reference measurement; the two parameters are selected separately (Sect.~\ref{sec:selection}), so the samples are nested but not identical. Columns are the median and the robust dispersion of the fractional residual $r=(\hat{y}-y)/y$, the fraction beyond $5\,\sigma_{\rm MAD}$ of the median, the fraction agreeing with the reference to better than $10\%$, and the fraction with $|R|<1$ for the normalised residual of Eq.~\ref{eq:norm-res}, which should equal $68.3\%$ if the reported uncertainties are well calibrated. The \emph{TESS} rows labelled \emph{all cross-matched} drop the reference reliability flag \texttt{f\_dnu}; see Sect.~\ref{sec:res-tess}.}
\end{table*}

\subsection{Held-out synthetic spectra}
\label{sec:res-synthetics}

The held-out spectra come from the same generator as the training set but from files the networks never saw. Because the reference is the value that was used to build the spectrum, the residual measures the error of the module directly, with no contribution from a reference pipeline.

Both modules are unbiased on this sample: the median fractional residual is below $0.02\%$ for $\nu_{\max}$ and for $\Delta\nu$. The dispersions are $\sigma_{\rm MAD}=1.38\%$ in $\nu_{\max}$ and $0.35\%$ in $\Delta\nu$. The residuals are flat across the trained range in both parameters (Figs.~\ref{fig:numax-comparison}a and~\ref{fig:dnu-comparison}a), with two exceptions. The $\nu_{\max}$ residuals turn slightly negative above $\sim250\,\mu$Hz, where the output bins stop and the distribution can no longer place probability beyond the edge; this is the behaviour that criterion (v) of Sect.~\ref{sec:selection} removes. The $\Delta\nu$ residuals broaden below $\Delta\nu \approx 2.5\,\mu$Hz, where the window around $\nu_{\max}$ contains the fewest resolution elements per radial order.

That $\Delta\nu$ is recovered four times more precisely than $\nu_{\max}$ is not surprising. $\Delta\nu$ is the spacing of a comb, and the position of a peak in the PSPS is set by the whole set of radial orders in the window, so the measurement averages over them. $\nu_{\max}$ is the centroid of a broad envelope whose individual mode heights are stochastic, and a single realisation of the $\chi^2_2$ noise displaces that centroid by an amount that no method can recover. A part of the $1.38\%$ is therefore an intrinsic property of the spectrum rather than a deficiency of the network.

The fraction of the sample lying beyond $5\,\sigma_{\rm MAD}$ is $6.5\%$ in $\nu_{\max}$, which is larger than for either observed sample and should not be read as a higher failure rate. The statistic compares the tails of the distribution with its core, and the core here is extremely narrow: five times $\sigma_{\rm MAD}$ is only $6.9\%$ in $\nu_{\max}$, so the modest shoulders visible in Fig.~\ref{fig:numax-comparison}d already exceed it. In absolute terms $98.1\%$ of the sample recovers $\nu_{\max}$ to better than $10\%$, and essentially the entire sample does so in $\Delta\nu$.

The composition of the synthetic sample contributes as well, since it spans a wider range of signal quality than either comparison set. The signal-to-noise ratio is sampled down to $5$, and the granulation amplitude and timescale are drawn over broad ranges (Table~\ref{tab:parameter_ranges}), so a fraction of the held-out spectra have an envelope that is weak or effectively buried in the background. No method would preform well for $\nu_{\max}$ in those spectra, and they are present by design, since the training set has to cover the faint end for the module to behave sensibly there. The \emph{Kepler} and \emph{TESS} comparison samples contain no equivalent population, because both are restricted to stars for which an earlier analysis had already reported a detection. Part of the synthetic tail is therefore a property of the test set rather than of the module.

Nonetheless, this sample provides the cleanest test of the reported uncertainties, since the reference carries none of its own and the normalised residual of Eq.~\ref{eq:norm-res} is formed from the output distribution alone. For $\nu_{\max}$ we find $65.4\%$ of the stars with $|R|<1$, against the $68.3\%$ expected for Gaussian errors. The median reported uncertainty, $1.30\%$, also matches the realised dispersion of $1.38\%$ closely. The $\nu_{\max}$ module therefore does not merely produce a point estimate with an error bar attached; the width of its output distribution is a quantitatively correct statement about how wrong that estimate is likely to be.

The $\Delta\nu$ module is calibrated differently. Its intervals are too wide: $87.8\%$ of the sample falls within $|R|<1$, and the median reported uncertainty of $0.60\%$ is nearly twice the realised dispersion of $0.35\%$. The explanation is likely the soft labels of Eq.~\ref{eq:soft-labels}. The $\nu_{\max}$ module is trained on one-hot targets and is free to concentrate its output in a single bin, whereas the $\Delta\nu$ module is trained on Gaussian targets of width $\sigma_b = 1$ bin. Since the bins are spaced by $0.5\%$ in $\log\Delta\nu$, no training example ever asks the network for a distribution narrower than about that, and the reported $0.60\%$ sits close to this floor while the information in the PSPS evidently supports $0.35\%$. The width of the soft label therefore trades a conservative uncertainty for a well-behaved loss on narrow bins. The $\Delta\nu$ intervals should be read accordingly, and a narrower $\sigma_b$ should be tried if sharper intervals are needed.

\subsection{Kepler spectra at the training resolution}
\label{sec:res-kepler}

The second test uses real stars while keeping the reference under control. We take the \emph{Kepler} red giants of Sect.~\ref{sec:data-kepler}, whose spectra are sampled on the one-year training grid, and compare the resulting predictions with \citet{Yu_2018}, who measured the same stars from the full four-year light curves. The reference is thus built from four times the frequency resolution of the input. Its own uncertainty is small compared with what a one-year spectrum, and the residual is, to a good approximation, the error of the framework on real data.

For $\nu_{\max}$ we find a median offset of $+0.81\%$ with $\sigma_{\rm MAD}=3.29\%$, and $0.28\%$ of the sample beyond $5\,\sigma_{\rm MAD}$; $97.3\%$ of the sample agrees with the reference to better than $10\%$. The dispersion is roughly $2.4$ times the synthetic value, which we attribute to real granulation backgrounds and real window functions. The small positive offset is harder to attribute. It is well below the dispersion and we do not regard it as significant for individual stars, but it is a systematic of the sample and should be allowed for when these values are used in a scaling relation. It may reflect the difference between the envelope centroid the network learns and the Gaussian fit to the background-subtracted spectrum that defines $\nu_{\max}$ in \citet{Yu_2018}; the two can differ when the background is steep.

For $\Delta\nu$ the agreement is considerably better, with a median offset of $+0.14\%$ and $\sigma_{\rm MAD}=0.61\%$. This is within a factor of two of the synthetic performance. The comb is a sharply defined feature, and the network locates it in real stars nearly as well as in simulated ones. Fig.~\ref{fig:dnu-comparison}b shows one localised departure, a modest excess of positive residuals between $\Delta\nu \approx 3.5$ and $4.5\,\mu$Hz. This is the red clump, where the mixed-mode spectrum is densest and the p-mode comb is most heavily contaminated.

The calibration seen on the synthetic sample survives the move to real data. For $\nu_{\max}$, $63.0\%$ of the stars have $|R|<1$, a little below the $68.3\%$ expected for Gaussian errors, and our median reported uncertainty of $2.36\%$ combined in quadrature with the $1.57\%$ quoted by \citet{Yu_2018} gives $2.83\%$ against an observed $3.29\%$. The framework is therefore mildly over-confident here, where on the synthetic sample it was not. The difference is small, and we attribute it to error sources absent from the synthetic spectra rather than to a defect of the output distribution: a network trained on simulated data can only quantify the uncertainty arising from the noise realisations it was shown. The $\Delta\nu$ intervals remain conservative, with $85.6\%$ within $|R|<1$, consistent with the soft-label floor discussed above.

\subsection{TESS red giants}
\label{sec:res-tess}

The third comparison is against \citet{Sreenivas_2026}, who measured $\nu_{\max}$ and $\Delta\nu$ for red giants in the \emph{TESS} continuous viewing zones. Their analysis uses the full baseline available for each star, a median of $690$ days against the one year we supply to the modules (Sect.~\ref{sec:data-refs}), so the reference is the more precise of the two but not overwhelmingly so. Part of the scatter is therefore contributed by the reference, and we cannot separate the two contributions. Since both analyses quote uncertainties, we can instead check whether the observed scatter is consistent with the uncertainties both pipelines quote.

The two samples should be defined symmetrically. We do not quote numbers for every spectrum we process; we apply the selection of Sect.~\ref{sec:selection} first. \citet{Sreenivas_2026} likewise publish a reliability flag, \texttt{f\_dnu}, which marks the $58\%$ of their $\Delta\nu$ measurements that pass their own vetting. Comparing our selected predictions against their full catalogue would compare samples selected on different criteria, so we take as our primary comparison the stars for which both analyses report a reliable value. The cut is defined by the reference authors, without reference to our predictions, and we use their least restrictive $\Delta\nu$ flag rather than the stricter gold sample. For completeness, Table~\ref{tab:validation} also lists the statistics against the full catalogue, and we discuss the difference below.

Of the $10\,073$ cross-matched stars with a flagged-reliable reference, our $\nu_{\max}$ selection retains $7981$, or $79\%$, and our $\Delta\nu$ selection $7220$, or $72\%$.

For $\nu_{\max}$ the median offset is $-0.22\%$ with $\sigma_{\rm MAD}=4.43\%$ and only $0.15\%$ of the sample beyond $5\,\sigma_{\rm MAD}$, the lowest catastrophic-failure rate of the three samples. Our median reported uncertainty here is $4.06\%$, against $2.36\%$ on \emph{Kepler}, an increase close to the factor by which the data quality degrades. Combined in quadrature with the $1.38\%$ median uncertainty quoted by \citet{Sreenivas_2026}, the two pipelines predict a scatter of $4.29\%$ where $4.43\%$ is observed. The agreement is therefore close to what both analyses claim, and we find no evidence that the modules fail on \emph{TESS} in a way that the lower signal-to-noise does not already account for. The residuals in Fig.~\ref{fig:numax-comparison}c widen towards low $\nu_{\max}$, where the envelope sits on the steeply rising granulation background and fewer radial orders are excited, and tighten to a few per cent above $100\,\mu$Hz.

For $\Delta\nu$ we find a median offset of $-0.16\%$ with $\sigma_{\rm MAD}=1.27\%$, and $99.6\%$ of the sample agrees with the reference to better than $10\%$. The quadrature sum of the two quoted uncertainties, $1.10\%$, again falls a little short of the observed $1.27\%$. Both parameters therefore behave on \emph{TESS} as they do on \emph{Kepler}, with the dispersion scaled up by roughly the factor the data quality implies and no new failure mode appearing.

Comparing against all $17\,198$ cross-matched stars the $\Delta\nu$ dispersion rises from $1.27\%$ to $1.84\%$, but the real difference is in the tails: $11.0\%$ of that sample lies beyond $5\,\sigma_{\rm MAD}$ against $1.2\%$ here, a factor of nine, and the excess is strongly one-sided. The calibration follows it down, from $65.2\%$ within $|R|<1$ to $55.3\%$. For $\nu_{\max}$, which the flag does not address directly, the degradation is milder: $4.86\%$ against $4.43\%$, and $0.89\%$ beyond $5\,\sigma_{\rm MAD}$ against $0.15\%$. Almost all of the discrepant stars carry $\texttt{f\_dnu}=0$, and inspecting their spectra shows our value sitting on a clear peak of the PSPS in almost all cases. The disagreement is restricted to measurements the reference catalogue flags as unreliable. It does, however, imply that the framework returns a usable $\Delta\nu$ for a population of stars missing from the earlier dataset, adding new inferences for stars in the \emph{TESS} continuous viewing zones.

The consistency check of Sect.~\ref{sec:selection} has a measurable effect here. The signal-to-noise module, used as the primary estimator, is immune to the low-frequency attenuation imposed by the five-day triangular detrending of Sect.~\ref{sec:data-tess}, which is why it is preferred for \emph{TESS}; but it is also the representation most exposed to a spectrum with no prominent envelope, where a single unrelated peak can dominate the prediction. Requiring agreement with the linear-input module removes those cases. The effect shows in the outlier fraction: at $0.15\%$ beyond $5\,\sigma_{\rm MAD}$, the \emph{TESS} $\nu_{\max}$ comparison has the lowest catastrophic-failure rate of the three samples despite being the noisiest data, because two representations must agree before a value is reported at all.

Across the three validation samples the criterion removes $4$ of the $16\,855$ synthetic spectra that would otherwise have been selected, $280$ of $14\,444$ on \emph{Kepler}, and $25$ of $8006$ on \emph{TESS}. The ordering is the opposite of what one might expect. The check does its heaviest work on the cleanest data, because the \emph{Kepler} spectra are the ones on which the signal-to-noise module is most liable to settle on a sharp unrelated feature and report it with a narrow, confident distribution, which is the failure the linear-input module is there to catch. On the \emph{TESS} cross-match the same criterion rejects only $0.3\%$, since a star with a reliable reference measurement is by construction one with a well-defined envelope. The spurious-signal test of Sect.~\ref{sec:spurious} removes a further $216$ \emph{TESS} stars from the selected set, having flagged $321$ in the cross-match as a whole. Neither figure carries over to the full sample of Sect.~\ref{sec:data-tess}, where both tests are far more active, and the cross-matched set is therefore a benign subsample, not a representative one.

The reported uncertainties are mildly over-confident on this sample, with $66.6\%$ of the $\nu_{\max}$ measurements and $65.2\%$ of the $\Delta\nu$ measurements inside $|R|<1$ against the expected $68.3\%$. The $\Delta\nu$ figure requires comment, because it is the only sample on which that module is not conservative. On the synthetic and \emph{Kepler} samples the soft-label floor discussed in Sect.~\ref{sec:res-synthetics} left the intervals too wide, at $87.8\%$ and $85.6\%$. On \emph{TESS} the additional error sources that the synthetic spectra do not contain are large enough to cancel that conservatism. The agreement with $68.3\%$ therefore arises from two effects cancelling, and we would not expect it to hold at a different signal-to-noise.

\subsection{Application to the TESS red giants}
\label{sec:res-catalogue}

We now apply the framework to the full sample of $274\,721$ stars of Sect.~\ref{sec:data-tess}. The selection of Sect.~\ref{sec:selection} returns a reliable $\nu_{\max}$ for $56\,224$ stars, or $20.5\%$ of the sample, and a reliable $\Delta\nu$ for $39\,939$, or $14.5\%$. A yield of this order is expected: the parent sample is a photometric selection of red giant candidates, not a list of confirmed oscillators, and a large fraction of it consists of stars too faint, too blended or too evolved for \emph{TESS} to resolve an envelope at all. The two vetting steps remove a small fraction of the sample. The spurious-signal test of Sect.~\ref{sec:spurious} flags $24\,959$ spectra in the full sample and removes $9\,417$ that would otherwise have been accepted, and the consistency check of Sect.~\ref{sec:selection} removes a further $1\,792$.

The cost of the analysis is small. On a single NVIDIA RTX A4000, the $\nu_{\max}$ module processes the full sample of $274\,721$ spectra in $2049$\,s and the $\Delta\nu$ module in $2228$\,s, which is $7.5$ and $8.1$\,ms per star. Both figures are end to end. They include reading each spectrum from disk, every preprocessing step, the cropping and the construction of the signal-to-noise and PSPS representations, and the interpretation of the output distribution into a median, quantiles, confidence, normalised entropy and the reliability index of Sect.~\ref{sec:uncertainties}. Characterising the whole sample therefore costs under two hours of wall-clock time on one card, counting the second pass of the $\nu_{\max}$ module that the consistency check of Sect.~\ref{sec:selection} requires, and the cost grows linearly with the number of stars. Only the periodograms are computed upstream of this, and they are a cost incurred once per star whatever the subsequent analysis. The hardware is modest, The relevant point is that no step of the procedure requires a person to inspect a spectrum. The sample size the method can handle is therefore set by the computing available.

Table~\ref{tab:catalogue} shows the first ten entries of the resulting catalogue, ordered by \emph{TESS} identifier. Four of the ten carry a $\nu_{\max}$ but no $\Delta\nu$, which is the ordinary situation rather than an exception: the two parameters are selected separately (Sect.~\ref{sec:selection}), and a star whose envelope is well enough defined to locate a power excess need not show a comb clean enough to measure its spacing. Anyone using the catalogue for $\nu_{\max}$ alone therefore has access to a sample $40\%$ larger than the one available for both parameters together.

% ------------------------------------------------------------------ catalogue
\begin{table}[htbp]
\caption{First ten entries of the \emph{TESS} catalogue.}
  \label{tab:catalogue}
  \centering
\setlength{\tabcolsep}{4pt}\small
  \begin{tabular}{r c c c c}
    \toprule
    & \multicolumn{2}{c}{$\nu_{\max}$ ($\mu$Hz)} &
      \multicolumn{2}{c}{$\Delta\nu$ ($\mu$Hz)} \\
    \cmidrule(lr){2-3}\cmidrule(lr){4-5}
    TIC & value & $\mathcal{R}$ & value & $\mathcal{R}$ \\
    \midrule
    $7547609$ & $29.45^{+0.94}_{-0.98}$ & $0.34$ & $4.14^{+0.03}_{-0.03}$ & $0.30$ \\[2pt]
    $7548894$ & $36.93^{+0.80}_{-0.76}$ & $0.35$ & $4.21^{+0.03}_{-0.03}$ & $0.23$ \\[2pt]
    $7581877$ & $10.97^{+0.79}_{-0.98}$ & $0.12$ & \ldots & \ldots \\[2pt]
    $7581996$ & $11.10^{+0.70}_{-0.88}$ & $0.17$ & \ldots & \ldots \\[2pt]
    $7582000$ & $11.08^{+0.68}_{-0.79}$ & $0.20$ & \ldots & \ldots \\[2pt]
    $7582118$ & $34.25^{+1.28}_{-1.17}$ & $0.24$ & $3.84^{+0.05}_{-0.04}$ & $0.16$ \\[2pt]
    $7582146$ & $8.64^{+0.76}_{-0.51}$  & $0.24$ & \ldots & \ldots \\[2pt]
    $7582335$ & $18.71^{+0.94}_{-0.77}$ & $0.26$ & $2.65^{+0.06}_{-0.06}$ & $0.14$ \\[2pt]
    $7582397$ & $22.61^{+1.02}_{-0.89}$ & $0.28$ & \ldots & \ldots \\[2pt]
    $7582528$ & $13.87^{+0.83}_{-0.73}$ & $0.22$ & $2.02^{+0.02}_{-0.02}$ & $0.19$ \\
    \bottomrule
  \end{tabular}
\tablefoot{Values are the medians of the predicted distributions and the errors the $16$th and $84$th percentiles, as defined in Sect.~\ref{sec:uncertainties}; $\mathcal{R}$ is the reliability index of Eq.~\ref{eq:ri}. An ellipsis marks a star for which $\nu_{\max}$ passes the selection of Sect.~\ref{sec:selection} but $\Delta\nu$ does not. The rows are the first ten of the catalogue by identifier. The published version additionally carries, for each parameter, the mode of the distribution, the confidence $\mathcal{C}$, the normalised entropy $\mathcal{H}$ and the individual selection flags, so that a user may impose a stricter cut than ours, depending on their science case. It will be made available in electronic form at the CDS.}
\end{table}

\begin{figure*}[htbp]
  \centering
  \includegraphics[width=0.7\linewidth]{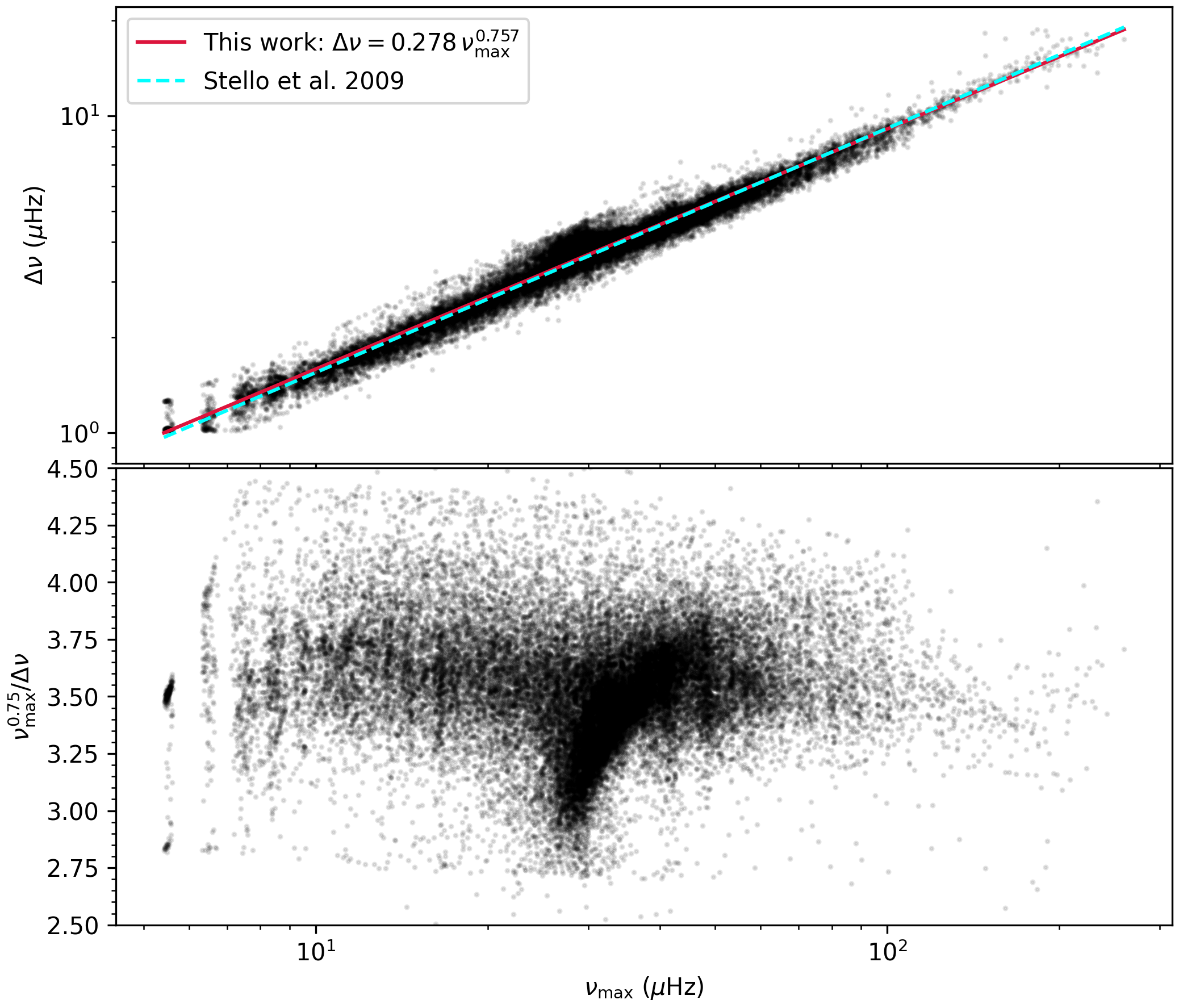}
\caption{Inferred $\Delta\nu$ against inferred $\nu_{\max}$ for the $39\,939$ \emph{TESS} red giants with a reliable measurement of both (\emph{top}), with the power law fitted to the sample (solid) and the relation of \citet{Stello_2009} (dashed) shown for comparison, and the ratio $\nu_{\max}^{0.75}/\Delta\nu$ against $\nu_{\max}$ (\emph{bottom}), which flattens the relation and expands its residual structure. The clear overdensity in the lower panel between $\nu_{\max} \approx 20$ and $40\,\mu$Hz, spreading downwards from the main ridge, corresponds to the red clump.}
  \label{fig:numax-dnu}
\end{figure*}

Figure~\ref{fig:numax-dnu} shows the two parameters against each other. A least-squares fit in logarithmic space over the $39\,939$ stars, spanning $5.4$--$260\,\mu$Hz, gives $\Delta\nu = 0.278\,\nu_{\max}^{0.757}$, with a robust dispersion of $6.2\%$ about the fit and a median residual of $-0.2\%$. The exponent sits close to the $0.772$ of \citet{Stello_2009}, and comparing the two exponents alone would understate how close: because the exponent and the normalisation trade off against one another, the relevant comparison is between the curves themselves. Ours runs $3.3\%$ above theirs at the bottom of the range, crosses it near $70\,\mu$Hz and falls $1.7\%$ below it at the top, so the two relations differ by less than the scatter of the sample everywhere. We note that our fit is made to a mixed population. The clump contributes a large share of the stars near $30\,\mu$Hz, and since that is essentially where the fit pivots, its main effect is on the normalisation rather than on the exponent; a sample weighted differently between the two evolutionary states would return a slightly different pair of numbers. This agreement should not be over-interpreted. It is not a test of the scaling relation, which is far better constrained by smaller and more precisely measured samples. It is a statement about the framework: the two modules are trained independently, on parameters that are sampled independently in the synthetic set (Sect.~\ref{sec:training-set}), and the selection contains no scaling relation of any kind. That the two inferences nonetheless fall on the expected locus, for forty thousand stars the networks have never seen, indicates that each is reading the structure it was meant to read rather than exploiting a correlation learned from the training set.

The lower panel shows a pronounced overdensity near $\nu_{\max} \approx 30\,\mu$Hz, where the points spread downwards from the main locus at $\nu_{\max}^{0.75}/\Delta\nu \approx 3.6$ to values near $2.8$. This is the red clump. Core-helium-burning stars at a given $\nu_{\max}$ have a larger $\Delta\nu$ than their shell-burning counterparts, because their interior structure differs, and they therefore sit above the relation in the upper panel and below the ridge in the lower one. Both evolutionary states are represented in the training set, but neither module is given an evolutionary label or asked to produce one, so the feature emerges from the measured parameters rather than from any classification. A secondary concentration is visible at higher $\nu_{\max}$ along the upper edge of the plume, in the region where the secondary clump is expected for more massive stars.

The two populations separate statistically but overlap for individual stars, and we do not attempt to classify individual objects here. A dedicated evolutionary-state classifier, operating on the spectrum in the same way as the modules described above, would separate them far more reliably than a cut in this plane, and could be added as a third module. We consider this a useful extension.

One artefact should be noted. At the lowest frequencies, below $\nu_{\max} \approx 7\,\mu$Hz, the points fall into narrow vertical stripes in both panels. These are not physical. They reflect the discreteness of the $\Delta\nu$ output bins of Sect.~\ref{sec:dnu-grid}, which are logarithmically spaced and therefore coarsest in absolute terms at the bottom of the range, combined with the small number of stars there. The affected stars lie at the edge of the trained range and carry correspondingly large uncertainties.

\subsection{The released catalogue}
\label{sec:release}

We release the $56\,224$ stars for which the framework returns a reliable $\nu_{\max}$, of which $39\,939$ also carry a reliable $\Delta\nu$. Each entry gives, for each parameter it has, the median and the $16$th and $84$th percentiles of the predicted distribution, its mode, the confidence $\mathcal{C}$, the normalised entropy $\mathcal{H}$ and the reliability index $\mathcal{R}$, together with the individual flags of Sect.~\ref{sec:selection} and the spurious-signal verdict of Appendix~\ref{app:spurious}. We do not publish the remainder of the $274\,721$ spectra that were processed. The parent sample is a photometric selection of candidates, and most of what the selection removes are stars too faint, too blended or too evolved to show an envelope at all, for which a catalogue entry would record nothing but a non-detection.

The diagnostics are released alongside the values, since the thresholds of Sect.~\ref{sec:selection} are the looser end of a range. The reliability index is constructed so that a prediction scores highly when its distribution is both confident and narrow relative to what the parameter can support. So a user can define the threshold depending on their science case, trading sample size against reliability. The catalogue will be made available in electronic form at the CDS.

\section{Summary and Conclusion}
\label{sec:conclusion}

We have presented ROSE, a modular machine-learning framework for the asteroseismic characterisation of red giants, and used two of its modules to infer $\nu_{\max}$ and $\Delta\nu$ from power spectra. A module is a single network together with the preprocessing transform that defines its input and the binning that defines its output, the three being stored as one object, so that modules can be retrained, replaced or added without disturbing anything else. The organising principle is that each parameter should be measured in the representation that displays it most directly: $\nu_{\max}$ from the full-range PSD, where the oscillation envelope is a broad excess of power, and $\Delta\nu$ from the PSPS of a window centred on $\nu_{\max}$, where the regular spacing of the pressure modes produces a family of harmonics of $\Delta\nu$, the full pattern constraining it more tightly than any single peak would.

Both modules are trained exclusively on synthetic spectra generated at a fixed one-year baseline, and the inference is posed as a classification over bins, so that each prediction is a probability distribution rather than a point estimate. Two consequences follow, and they are the practical case for the design. The uncertainty is read from the width of that distribution rather than assumed, and its shape supplies the confidence, the normalised entropy and the reliability index on which the selection of the reported sample rests. The second is that the framework never uses a scaling relation to arrive at an answer. The one place a relation appears at all is in generating the training set, where $\nu_{\max}$ is drawn with a $30\%$ spread about the value implied by $\Delta\nu$, broad enough that a network leaning on the correlation is penalised rather than rewarded. At inference nothing of the kind is used: the $\Delta\nu$ module takes $\nu_{\max}$ only to choose a window, never as a predictor of its own answer, and no relation enters the selection.

Our main results are as follows.

\begin{enumerate}
  \item On held-out synthetic spectra both modules are unbiased, with robust dispersions of $1.38\%$ in $\nu_{\max}$ and $0.35\%$ in $\Delta\nu$. On \emph{Kepler} spectra sampled at the one-year training resolution and compared with \citet{Yu_2018}, the dispersions are $3.29\%$ and $0.61\%$; on \emph{TESS} spectra compared with \citet{Sreenivas_2026}, $4.43\%$ and $1.27\%$. The degradation from synthetic to real data is modest, and the framework performs well on observations.

  \item The reported $\nu_{\max}$ uncertainties are close to their nominal coverage on all three samples, with $63$--$67\%$ of stars inside $|R|<1$ against the $68.3\%$ expected. The $\Delta\nu$ intervals are conservative by roughly a factor of two, which we attribute to the width of the soft labels used in training and not to the information content of the PSPS.
  \item Applied to $274\,721$ \emph{TESS} red giant candidates observed in at least ten sectors, the framework returns a reliable $\nu_{\max}$ for $56\,224$ stars and a reliable $\Delta\nu$ for $39\,939$. The two samples are quoted separately, since a catalogue used for $\nu_{\max}$ alone should not be limited by the completeness of a parameter it does not use.
  \item The resulting sample follows $\Delta\nu = 0.278\,\nu_{\max}^{0.757}$ with a scatter of $6.2\%$, within $3.3\%$ of \citet{Stello_2009} everywhere over $5.4$--$260\,\mu$Hz. The red clump appears as a clear overdensity in the $\nu_{\max}^{0.75}/\Delta\nu$ plane. Both evolutionary states are represented in the training set, but neither module is given an evolutionary label or asked to produce one, so the feature emerges from the measured parameters rather than from any classification.
  \item The $56\,224$ stars with a reliable $\nu_{\max}$ are released, together with the predicted distributions, their summary statistics and every diagnostic entering the selection, so that users may impose a stricter threshold than ours and trade sample size against reliability as their science requires.
  \item The analysis costs $7.5$ and $8.1$\,ms per star for the two modules, end to end, so the full sample is characterised in under two hours on a single GPU, with no manual inspection.
\end{enumerate}

The framework has three principal limitations, each of which suggests a natural extension. First, the $\Delta\nu$ module is conditioned on $\nu_{\max}$ and therefore inherits its failures, which is why the two selections are nested. Providing the two $\nu_{\max}$ representations to a single network as parallel inputs, rather than using separate models, may allow the network to exploit complementary information from both representations, potentially improving performance while eliminating the need for a separate consistency check. Second, the spurious-signal criteria of Appendix~\ref{app:spurious} are empirical, tuned on \emph{TESS} spectra at a one-year resolution, and should not be applied to other data without re-examination. A trained classifier would provide a more robust replacement and could be incorporated as an additional module. Third, an evolutionary-state classifier is needed to distinguish giant-branch stars from core-helium-burning clump stars. Synthetic spectra provide a natural source of training data for this task, while existing verified samples can be retained for independent benchmarking. Finally, ongoing work is extending the framework to mixed-mode observables, including $\Delta\Pi_1$ and $q$, which would provide access to the stellar core and substantially expand the range of physical parameters that can be inferred.

These extensions are particularly timely given the rapidly expanding volume of asteroseismic data. TESS has already produced a large stellar sample, and PLATO and Roman will increase both the scale and diversity of available observations. Scalable, fast, and automated inference methods will therefore be essential for converting these datasets into precise and physically informative stellar measurements.

\begin{acknowledgments}
\nolinenumbers
The authors would like to thank Bastien Liagre and Savita Mathur for useful comments and discussions. N.G. and S.H. acknowledge support from the Department of Atomic Energy, Government of India, under Project Identification No. RTI 4002. D.P. and R.A.G. acknowledge financial support from the Centre national d'\'etudes spatiales (CNES), France (ROR: \url{https://ror.org/04h1h0y33}), within the framework of the GOLF/SOHO and PLATO space missions. This paper includes data collected by the \emph{Kepler} and \emph{TESS} missions and obtained from the MAST data archive at the Space Telescope Science Institute (STScI). Funding to US Institutions for the \emph{Kepler} mission was provided by the NASA Science Mission Directorate. STScI is operated by the Association of Universities for Research in Astronomy, Inc., under NASA contract NAS 5-26555. This research was supported in part by a generous donation (from the Murty Trust) aimed at enabling advances in astrophysics through the use of machine learning. Murty Trust, an initiative of the Murty Foundation, is a not-for-profit organisation dedicated to the preservation and celebration of culture, science, and knowledge systems born out of India. The Murty Trust is headed by Mrs. Sudha Murty and Mr. Rohan Murty.

We also acknowledge the use of Claude, developed by Anthropic, for assistance in language editing during the preparation of this manuscript.
\end{acknowledgments}

\facilities{Kepler, TESS}
\software{NumPy \citep{harris2020array},
Lightkurve \citep{Lightkurve_2018}, TensorFlow \citep{tensorflow2015-whitepaper}, Pandas \citep{reback_2022_6702671}}

\clearpage
\appendix
\section{Spurious-signal criteria}
\label{app:spurious}

This appendix collects the detail of the morphological test described in Sect.~\ref{sec:spurious}. As noted there, the thresholds are empirical and tuned for \emph{TESS} spectra at a one-year resolution.

For each star the PSD is normalised by its maximum and flattened with the same sliding logarithmic median filter as in Sect.~\ref{sec:psps}, giving a signal-to-noise spectrum. We locate its peaks and retain those whose prominence is at least half their height, which excludes the noise fluctuations riding on the flank of a broader structure, and keep the ten tallest. Peaks within three samples of a taller one are merged, since an unresolved coherent signal leaks across adjacent bins and would otherwise be counted more than once. We denote the sorted heights $h_1 \ge h_2 \ge \dots$ and the frequency of the tallest peak $f_1$, and define the concentration $\mathcal{C}$ as the fraction of the total retained peak height lying within $\pm0.75\,\delta\nu_{\rm env}(f_1)$ of $f_1$, with $\delta\nu_{\rm env}$ from Eq.~\ref{eq:fwhm}.

Five tests are applied, each of which identifies a configuration that a stochastically excited Gaussian envelope cannot produce:
\begin{itemize}
  \item $h_1 > 100$. A stochastic mode in a background-divided spectrum does not reach this height; the peak is coherent.
  \item $h_1 > 40$ and $h_2/h_1 < 0.25$. The envelope places the adjacent radial order close to $h_1$, so a second peak below a quarter of the first indicates an isolated spike rather than an envelope.
  \item $h_1 > 40$, $h_3/h_1 > 0.7$ and $\mathcal{C} < 0.35$. Three peaks of comparable height belonging to one envelope would have to be concentrated within it; comparable heights combined with a low concentration cannot arise from a Gaussian envelope.
  \item $h_1 > 15$ and at least three peaks above the noise floor lying within $5\%$ of $f_1$ of an integer multiple of $f_1$ or $f_1/2$. Pressure modes are spaced additively by $\Delta\nu$ whereas harmonics are spaced multiplicatively; the two combs are incompatible.
  \item $h_1 > 30$, at least three peaks above a quarter of $h_1$ lying outside the envelope window, and a span of comparable peaks exceeding $2.5\,\delta\nu_{\rm env}(f_1)$. An envelope of this width cannot contain peaks distributed over this range.
\end{itemize}
A spectrum triggering any of the five is flagged as spurious. A spectrum whose tallest peak satisfies $h_1 < 12$ is flagged as indeterminate. The remainder are flagged as clean.

\bibliography{ref}

\begin{thebibliography}{}
\expandafter\ifx\csname natexlab\endcsname\relax\def\natexlab#1{#1}\fi
\providecommand{\url}[1]{\href{#1}{#1}}
\providecommand{\dodoi}[1]{doi:~\href{http://doi.org/#1}{\nolinkurl{#1}}}
\providecommand{\doeprint}[1]{\href{http://ascl.net/#1}{\nolinkurl{http://ascl.net/#1}}}
\providecommand{\doarXiv}[1]{\href{https://arxiv.org/abs/#1}{\nolinkurl{https://arxiv.org/abs/#1}}}

\bibitem[{Abadi {et~al.}(2015)Abadi, Agarwal, Barham, Brevdo, Chen, Citro, Corrado, Davis, Dean, Devin, Ghemawat, Goodfellow, Harp, Irving, Isard, Jia, Jozefowicz, Kaiser, Kudlur, Levenberg, Man\'{e}, Monga, Moore, Murray, Olah, Schuster, Shlens, Steiner, Sutskever, Talwar, Tucker, Vanhoucke, Vasudevan, Vi\'{e}gas, Vinyals, Warden, Wattenberg, Wicke, Yu, \& Zheng}]{tensorflow2015-whitepaper}
Abadi, M., Agarwal, A., Barham, P., {et~al.} 2015.
\newblock \url{https://www.usenix.org/system/files/conference/osdi16/osdi16-abadi.pdf}

\bibitem[{{Anders} {et~al.}(2017){Anders}, {Chiappini}, {Minchev}, {Miglio}, {Montalb{\'a}n}, {Mosser}, {Rodrigues}, {Santiago}, {Baudin}, {Beers}, {da Costa}, {Garc{\'\i}a}, {Garc{\'\i}a-Hern{\'a}ndez}, {Holtzman}, {Maia}, {Majewski}, {Mathur}, {Noels-Grotsch}, {Pan}, {Schneider}, {Schultheis}, {Steinmetz}, {Valentini}, \& {Zamora}}]{Anders_2017}
{Anders}, F., {Chiappini}, C., {Minchev}, I., {et~al.} 2017, \aap, 600, A70, \dodoi{10.1051/0004-6361/201629363}

\bibitem[{{Beck} {et~al.}(2026){Beck}, {Masseron}, {Pavlovski}, {Godoy-Rivera}, {Mathur}, {Grossmann}, {Hamy}, {Palakkatharappil}, {Panetier}, {Garc{\'\i}a}, {Merc}, {Lu}, {Amestoy}, \& {Deeg}}]{Beck_2026}
{Beck}, P.~G., {Masseron}, T., {Pavlovski}, K., {et~al.} 2026, \aap, 706, L19, \dodoi{10.1051/0004-6361/202557452}

\bibitem[{Benomar(2023)}]{othman_benomar_2023_spectra_simulator}
Benomar, O. 2023, Power Spectrum Simulator for Solar-like stars, Siddharth2023,  Zenodo, \dodoi{10.5281/zenodo.8296459}

\bibitem[{Bishop(1994)}]{bishop1994_mixture_density_networks}
Bishop, C. 1994, Mixture density networks, WorkingPaper 4288, Aston University

\bibitem[{{Borucki} {et~al.}(2010){Borucki}, {Koch}, {Basri}, {Batalha}, {Brown}, {Caldwell}, {Caldwell}, {Christensen-Dalsgaard}, {Cochran}, {DeVore}, {Dunham}, {Dupree}, {Gautier}, {Geary}, {Gilliland}, {Gould}, {Howell}, {Jenkins}, {Kondo}, {Latham}, {Marcy}, {Meibom}, {Kjeldsen}, {Lissauer}, {Monet}, {Morrison}, {Sasselov}, {Tarter}, {Boss}, {Brownlee}, {Owen}, {Buzasi}, {Charbonneau}, {Doyle}, {Fortney}, {Ford}, {Holman}, {Seager}, {Steffen}, {Welsh}, {Rowe}, {Anderson}, {Buchhave}, {Ciardi}, {Walkowicz}, {Sherry}, {Horch}, {Isaacson}, {Everett}, {Fischer}, {Torres}, {Johnson}, {Endl}, {MacQueen}, {Bryson}, {Dotson}, {Haas}, {Kolodziejczak}, {Van Cleve}, {Chandrasekaran}, {Twicken}, {Quintana}, {Clarke}, {Allen}, {Li}, {Wu}, {Tenenbaum}, {Verner}, {Bruhweiler}, {Barnes}, \& {Prsa}}]{Kepler_mission}
{Borucki}, W.~J., {Koch}, D., {Basri}, G., {et~al.} 2010, Science, 327, 977, \dodoi{10.1126/science.1185402}

\bibitem[{Davies \& Miglio(2016)}]{davies2016AsteroseismologyRedGiantsa}
Davies, G.~R., \& Miglio, A. 2016, Astronomische Nachrichten, 337, 774, \dodoi{10.1002/asna.201612371}

\bibitem[{{Dhanpal} {et~al.}(2022){Dhanpal}, {Benomar}, {Hanasoge}, {Kundu}, {Dhuri}, {Das}, \& {Kaul}}]{dhanpal2022}
{Dhanpal}, S., {Benomar}, O., {Hanasoge}, S., {et~al.} 2022, \apj, 928, 188, \dodoi{10.3847/1538-4357/ac5247}

\bibitem[{{Dhanpal} {et~al.}(2023){Dhanpal}, {Benomar}, {Hanasoge}, {Takata}, {Panda}, \& {Kundu}}]{dhanpal2023}
---. 2023, \apj, 958, 63, \dodoi{10.3847/1538-4357/ad0046}

\bibitem[{{Dupret} {et~al.}(2009){Dupret}, {Belkacem}, {Samadi}, {Montalban}, {Moreira}, {Miglio}, {Godart}, {Ventura}, {Ludwig}, {Grigahc{\`e}ne}, {Goupil}, {Noels}, \& {Caffau}}]{Dupret_2009}
{Dupret}, M.~A., {Belkacem}, K., {Samadi}, R., {et~al.} 2009, \aap, 506, 57, \dodoi{10.1051/0004-6361/200911713}

\bibitem[{{Gaia Collaboration} {et~al.}(2018){Gaia Collaboration}, {Brown}, {Vallenari}, {Prusti}, {de Bruijne}, {Babusiaux}, {Bailer-Jones}, {Biermann}, {Evans}, {Eyer}, {Jansen}, {Jordi}, {Klioner}, {Lammers}, {Lindegren}, {Luri}, {Mignard}, {Panem}, {Pourbaix}, {Randich}, {Sartoretti}, {Siddiqui}, {Soubiran}, {van Leeuwen}, {Walton}, {Arenou}, {Bastian}, {Cropper}, {Drimmel}, {Katz}, {Lattanzi}, {Bakker}, {Cacciari}, {Casta{\~n}eda}, {Chaoul}, {Cheek}, {De Angeli}, {Fabricius}, {Guerra}, {Holl}, {Masana}, {Messineo}, {Mowlavi}, {Nienartowicz}, {Panuzzo}, {Portell}, {Riello}, {Seabroke}, {Tanga}, {Th{\'e}venin}, {Gracia-Abril}, {Comoretto}, {Garcia-Reinaldos}, {Teyssier}, {Altmann}, {Andrae}, {Audard}, {Bellas-Velidis}, {Benson}, {Berthier}, {Blomme}, {Burgess}, {Busso}, {Carry}, {Cellino}, {Clementini}, {Clotet}, {Creevey}, {Davidson}, {De Ridder}, {Delchambre}, {Dell'Oro}, {Ducourant}, {Fern{\'a}ndez-Hern{\'a}ndez}, {Fouesneau}, {Fr{\'e}mat}, {Galluccio}, {Garc{\'\i}a-Torres}, {Gonz{\'a}lez-N{\'u}{\~n}ez}, {Gonz{\'a}lez-Vidal}, {Gosset}, {Guy}, {Halbwachs}, {Hambly}, {Harrison}, {Hern{\'a}ndez}, {Hestroffer}, {Hodgkin}, {Hutton}, {Jasniewicz}, {Jean-Antoine-Piccolo}, {Jordan}, {Korn}, {Krone-Martins}, {Lanzafame}, {Lebzelter}, {L{\"o}ffler}, {Manteiga}, {Marrese}, {Mart{\'\i}n-Fleitas}, {Moitinho}, {Mora}, {Muinonen}, {Osinde}, {Pancino}, {Pauwels}, {Petit}, {Recio-Blanco}, {Richards}, {Rimoldini}, {Robin}, {Sarro}, {Siopis}, {Smith}, {Sozzetti}, {S{\"u}veges}, {Torra}, {van Reeven}, {Abbas}, {Abreu Aramburu}, {Accart}, {Aerts}, {Altavilla}, {{\'A}lvarez}, {Alvarez}, {Alves}, {Anderson}, {Andrei}, {Anglada Varela}, {Antiche}, {Antoja}, {Arcay}, {Astraatmadja}, {Bach}, {Baker}, {Balaguer-N{\'u}{\~n}ez}, {Balm}, {Barache}, {Barata}, {Barbato}, {Barblan}, {Barklem}, {Barrado}, {Barros}, {Barstow}, {Bartholom{\'e} Mu{\~n}oz}, {Bassilana}, {Becciani}, {Bellazzini}, {Berihuete}, {Bertone}, {Bianchi}, {Bienaym{\'e}}, {Blanco-Cuaresma}, {Boch}, {Boeche}, {Bombrun}, {Borrachero}, {Bossini}, {Bouquillon}, {Bourda}, {Bragaglia}, {Bramante}, {Breddels}, {Bressan}, {Brouillet}, {Br{\"u}semeister}, {Brugaletta}, {Bucciarelli}, {Burlacu}, {Busonero}, {Butkevich}, {Buzzi}, {Caffau}, {Cancelliere}, {Cannizzaro}, {Cantat-Gaudin}, {Carballo}, {Carlucci}, {Carrasco}, {Casamiquela}, {Castellani}, {Castro-Ginard}, {Charlot}, {Chemin}, {Chiavassa}, {Cocozza}, {Costigan}, {Cowell}, {Crifo}, {Crosta}, {Crowley}, {Cuypers}, {Dafonte}, {Damerdji}, {Dapergolas}, {David}, {David}, {de Laverny}, {De Luise}, {De March}, {de Martino}, {de Souza}, {de Torres}, {Debosscher}, {del Pozo}, {Delbo}, {Delgado}, {Delgado}, {Di Matteo}, {Diakite}, {Diener}, {Distefano}, {Dolding}, {Drazinos}, {Dur{\'a}n}, {Edvardsson}, {Enke}, {Eriksson}, {Esquej}, {Eynard Bontemps}, {Fabre}, {Fabrizio}, {Faigler}, {Falc{\~a}o}, {Farr{\`a}s Casas}, {Federici}, {Fedorets}, {Fernique}, {Figueras}, {Filippi}, {Findeisen}, {Fonti}, {Fraile}, {Fraser}, {Fr{\'e}zouls}, {Gai}, {Galleti}, {Garabato}, {Garc{\'\i}a-Sedano}, {Garofalo}, {Garralda}, {Gavel}, {Gavras}, {Gerssen}, {Geyer}, {Giacobbe}, {Gilmore}, {Girona}, {Giuffrida}, {Glass}, {Gomes}, {Granvik}, {Gueguen}, {Guerrier}, {Guiraud}, {Guti{\'e}rrez-S{\'a}nchez}, {Haigron}, {Hatzidimitriou}, {Hauser}, {Haywood}, {Heiter}, {Helmi}, {Heu}, {Hilger}, {Hobbs}, {Hofmann}, {Holland}, {Huckle}, {Hypki}, {Icardi}, {Jan{\ss}en}, {Jevardat de Fombelle}, {Jonker}, {Juh{\'a}sz}, {Julbe}, {Karampelas}, {Kewley}, {Klar}, {Kochoska}, {Kohley}, {Kolenberg}, {Kontizas}, {Kontizas}, {Koposov}, {Kordopatis}, {Kostrzewa-Rutkowska}, {Koubsky}, {Lambert}, {Lanza}, {Lasne}, {Lavigne}, {Le Fustec}, {Le Poncin-Lafitte}, {Lebreton}, {Leccia}, {Leclerc}, {Lecoeur-Taibi}, {Lenhardt}, {Leroux}, {Liao}, {Licata}, {Lindstr{\o}m}, {Lister}, {Livanou}, {Lobel}, {L{\'o}pez}, {Managau}, {Mann}, {Mantelet}, {Marchal}, {Marchant}, {Marconi}, {Marinoni}, {Marschalk{\'o}}, {Marshall}, {Martino}, {Marton}, {Mary}, {Massari}, {Matijevi{\v{c}}}, {Mazeh}, {McMillan}, {Messina}, {Michalik}, {Millar}, {Molina}, {Molinaro}, {Moln{\'a}r}, {Montegriffo}, {Mor}, {Morbidelli}, {Morel}, {Morris}, {Mulone}, {Muraveva}, {Musella}, {Nelemans}, {Nicastro}, {Noval}, {O'Mullane}, {Ord{\'e}novic}, {Ord{\'o}{\~n}ez-Blanco}, {Osborne}, {Pagani}, {Pagano}, {Pailler}, {Palacin}, {Palaversa}, {Panahi}, {Pawlak}, {Piersimoni}, {Pineau}, {Plachy}, {Plum}, {Poggio}, {Poujoulet}, {Pr{\v{s}}a}, {Pulone}, {Racero}, {Ragaini}, {Rambaux}, {Ramos-Lerate}, {Regibo}, {Reyl{\'e}}, {Riclet}, {Ripepi}, {Riva}, {Rivard}, {Rixon}, {Roegiers}, {Roelens}, {Romero-G{\'o}mez}, {Rowell}, {Royer}, {Ruiz-Dern}, {Sadowski}, {Sagrist{\`a} Sell{\'e}s}, {Sahlmann}, {Salgado}, {Salguero}, {Sanna}, {Santana-Ros}, {Sarasso}, {Savietto}, {Schultheis}, {Sciacca}, {Segol}, {Segovia}, {S{\'e}gransan}, {Shih}, {Siltala}, {Silva}, {Smart}, {Smith}, {Solano}, {Solitro}, {Sordo}, {Soria Nieto}, {Souchay}, {Spagna}, {Spoto}, {Stampa}, {Steele}, {Steidelm{\"u}ller}, {Stephenson}, {Stoev}, {Suess}, {Surdej}, {Szabados}, {Szegedi-Elek}, {Tapiador}, {Taris}, {Tauran}, {Taylor}, {Teixeira}, {Terrett}, {Teyssandier}, {Thuillot}, {Titarenko}, {Torra Clotet}, {Turon}, {Ulla}, {Utrilla}, {Uzzi}, {Vaillant}, {Valentini}, {Valette}, {van Elteren}, {Van Hemelryck}, {van Leeuwen}, {Vaschetto}, {Vecchiato}, {Veljanoski}, {Viala}, {Vicente}, {Vogt}, {von Essen}, {Voss}, {Votruba}, {Voutsinas}, {Walmsley}, {Weiler}, {Wertz}, {Wevers}, {Wyrzykowski}, {Yoldas}, {{\v{Z}}erjal}, {Ziaeepour}, {Zorec}, {Zschocke}, {Zucker}, {Zurbach}, \& {Zwitter}}]{Gaia_DR2_2018A&A...616A...1G}
{Gaia Collaboration}, {Brown}, A.~G.~A., {Vallenari}, A., {et~al.} 2018, \aap, 616, A1, \dodoi{10.1051/0004-6361/201833051}

\bibitem[{{Garc{\'\i}a} \& {Ballot}(2019)}]{Garcia_2019_review}
{Garc{\'\i}a}, R.~A., \& {Ballot}, J. 2019, Living Reviews in Solar Physics, 16, 4, \dodoi{10.1007/s41116-019-0020-1}

\bibitem[{{Garc{\'\i}a} {et~al.}(2011){Garc{\'\i}a}, {Hekker}, {Stello}, {Gutierrez-Soto}, {Handberg}, {Huber}, {Karoff}, {Uytterhoeven}, {Appourchaux}, {Chaplin}, {Elsworth}, {Mathur}, {Ballot}, {Christensen-Dalsgaard}, {Gilliland}, {Houdek}, {Jenkins}, {Kjeldsen}, {McCauliff}, {Metcalfe}, {Middour}, {Molenda-{\.Z}akowicz}, {Monteiro}, {Smith}, \& {Thompson}}]{Garcia_2011_KADACS}
{Garc{\'\i}a}, R.~A., {Hekker}, S., {Stello}, D., {et~al.} 2011, \mnras, 414, L6, \dodoi{10.1111/j.1745-3933.2011.01042.x}

\bibitem[{{Garc{\'\i}a} {et~al.}(2014){Garc{\'\i}a}, {Mathur}, {Pires}, {R{\'e}gulo}, {Bellamy}, {Pall{\'e}}, {Ballot}, {Barcel{\'o} Forteza}, {Beck}, {Bedding}, {Ceillier}, {Roca Cort{\'e}s}, {Salabert}, \& {Stello}}]{garcia2014ImpactGapsKepler}
{Garc{\'\i}a}, R.~A., {Mathur}, S., {Pires}, S., {et~al.} 2014, \aap, 568, A10, \dodoi{10.1051/0004-6361/201323326}

\bibitem[{{Garc{\'\i}a} {et~al.}(2024{\natexlab{a}}){Garc{\'\i}a}, {Palakkatharappil}, {Mathur}, {Pinsonneault}, {Ash}, {Breton}, {Bugnet}, {Cao}, {Gilles}, {Godoy-Rivera}, {Hamy}, {Liagre}, {Patton}, {Prin}, {Santos}, {Beck}, \& {Serenelli}}]{garcia2024measuringrotationperiods}
{Garc{\'\i}a}, R.~A., {Palakkatharappil}, D.~B., {Mathur}, S., {et~al.} 2024{\natexlab{a}}, in TESS Science Conference III, 32, \dodoi{10.5281/zenodo.13650178}

\bibitem[{{Garc{\'\i}a} {et~al.}(2024{\natexlab{b}}){Garc{\'\i}a}, {Palakkatharappil}, {Bugnet}, {Mathur}, {Liagre}, {Serenelli}, {Godoy-Rivera}, {Hamy}, {Prin}, {Beck}, {Pinsonneault}, \& {Davies}}]{garcia2024massivecharacterisation}
{Garc{\'\i}a}, R.~A., {Palakkatharappil}, D.~B., {Bugnet}, L., {et~al.} 2024{\natexlab{b}}, in 8th TESS/15th Kepler Asteroseismic Science Consortium Workshop, ed. T.~L. {Campante} (Zenodo), \dodoi{10.5281/zenodo.13647412}

\bibitem[{Ghanghas {et~al.}(2026)Ghanghas, Dhanpal, Hanasoge, Netrapalli, \& Shanmugam}]{ghanghas2026periodspacingsglobalseismic}
Ghanghas, N., Dhanpal, S., Hanasoge, S., Netrapalli, P., \& Shanmugam, K. 2026, Period spacings and global seismic parameters for K2 red giants using deep learning.
\newblock \doarXiv{2605.08051}

\bibitem[{{Goupil} {et~al.}(2013){Goupil}, {Mosser}, {Marques}, {Ouazzani}, {Belkacem}, {Lebreton}, \& {Samadi}}]{Goupil_2013}
{Goupil}, M.~J., {Mosser}, B., {Marques}, J.~P., {et~al.} 2013, \aap, 549, A75, \dodoi{10.1051/0004-6361/201220266}

\bibitem[{{Grosjean} {et~al.}(2014){Grosjean}, {Dupret}, {Belkacem}, {Montalban}, {Samadi}, \& {Mosser}}]{Grosjean_2014A&A...572A..11G}
{Grosjean}, M., {Dupret}, M.~A., {Belkacem}, K., {et~al.} 2014, \aap, 572, A11, \dodoi{10.1051/0004-6361/201423827}

\bibitem[{Harris {et~al.}(2020)Harris, Millman, van~der Walt, Gommers, Virtanen, Cournapeau, Wieser, Taylor, Berg, Smith, Kern, Picus, Hoyer, van Kerkwijk, Brett, Haldane, del R{\'{i}}o, Wiebe, Peterson, G{\'{e}}rard-Marchant, Sheppard, Reddy, Weckesser, Abbasi, Gohlke, \& Oliphant}]{harris2020array}
Harris, C.~R., Millman, K.~J., van~der Walt, S.~J., {et~al.} 2020, Nature, 585, 357, \dodoi{10.1038/s41586-020-2649-2}

\bibitem[{{Hekker} {et~al.}(2010){Hekker}, {Broomhall}, {Chaplin}, {Elsworth}, {Fletcher}, {New}, {Arentoft}, {Quirion}, \& {Kjeldsen}}]{Hekker_2010_BHM}
{Hekker}, S., {Broomhall}, A.-M., {Chaplin}, W.~J., {et~al.} 2010, \mnras, 402, 2049, \dodoi{10.1111/j.1365-2966.2009.16030.x}

\bibitem[{{Hon} {et~al.}(2022){Hon}, {Kuszlewicz}, {Huber}, {Stello}, \& {Reyes}}]{Hon_2022AJ....164..135H}
{Hon}, M., {Kuszlewicz}, J.~S., {Huber}, D., {Stello}, D., \& {Reyes}, C. 2022, \aj, 164, 135, \dodoi{10.3847/1538-3881/ac8931}

\bibitem[{{Hon} {et~al.}(2019){Hon}, {Stello}, {Garc{\'\i}a}, {Mathur}, {Sharma}, {Colman}, \& {Bugnet}}]{hon_kepler_giant_detections}
{Hon}, M., {Stello}, D., {Garc{\'\i}a}, R.~A., {et~al.} 2019, \mnras, 485, 5616, \dodoi{10.1093/mnras/stz622}

\bibitem[{{Hon} {et~al.}(2017){Hon}, {Stello}, \& {Yu}}]{Hon2017}
{Hon}, M., {Stello}, D., \& {Yu}, J. 2017, \mnras, 469, 4578, \dodoi{10.1093/mnras/stx1174}

\bibitem[{{Hon} {et~al.}(2018){Hon}, {Stello}, \& {Zinn}}]{Hon_2018}
{Hon}, M., {Stello}, D., \& {Zinn}, J.~C. 2018, \apj, 859, 64, \dodoi{10.3847/1538-4357/aabfdb}

\bibitem[{{Hon} {et~al.}(2021){Hon}, {Huber}, {Kuszlewicz}, {Stello}, {Sharma}, {Tayar}, {Zinn}, {Vrard}, \& {Pinsonneault}}]{quick_look_Hon_el_al}
{Hon}, M., {Huber}, D., {Kuszlewicz}, J.~S., {et~al.} 2021, \apj, 919, 131, \dodoi{10.3847/1538-4357/ac14b1}

\bibitem[{{Huang} {et~al.}(2020{\natexlab{a}}){Huang}, {Vanderburg}, {P{\'a}l}, {Sha}, {Yu}, {Fong}, {Fausnaugh}, {Shporer}, {Guerrero}, {Vanderspek}, \& {Ricker}}]{Huang_2020_QLP_I}
{Huang}, C.~X., {Vanderburg}, A., {P{\'a}l}, A., {et~al.} 2020{\natexlab{a}}, Research Notes of the American Astronomical Society, 4, 204, \dodoi{10.3847/2515-5172/abca2e}

\bibitem[{{Huang} {et~al.}(2020{\natexlab{b}}){Huang}, {Vanderburg}, {P{\'a}l}, {Sha}, {Yu}, {Fong}, {Fausnaugh}, {Shporer}, {Guerrero}, {Vanderspek}, \& {Ricker}}]{Huang_2020_QLP_II}
---. 2020{\natexlab{b}}, Research Notes of the American Astronomical Society, 4, 206, \dodoi{10.3847/2515-5172/abca2d}

\bibitem[{{Huber} {et~al.}(2009){Huber}, {Stello}, {Bedding}, {Chaplin}, {Arentoft}, {Quirion}, \& {Kjeldsen}}]{Huber_2009_SYD}
{Huber}, D., {Stello}, D., {Bedding}, T.~R., {et~al.} 2009, Communications in Asteroseismology, 160, 74, \dodoi{10.48550/arXiv.0910.2764}

\bibitem[{{Huber} {et~al.}(2011){Huber}, {Bedding}, {Stello}, {Hekker}, {Mathur}, {Mosser}, {Verner}, {Bonanno}, {Buzasi}, {Campante}, {Elsworth}, {Hale}, {Kallinger}, {Silva Aguirre}, {Chaplin}, {De Ridder}, {Garc{\'\i}a}, {Appourchaux}, {Frandsen}, {Houdek}, {Molenda-{\.Z}akowicz}, {Monteiro}, {Christensen-Dalsgaard}, {Gilliland}, {Kawaler}, {Kjeldsen}, {Broomhall}, {Corsaro}, {Salabert}, {Sanderfer}, {Seader}, \& {Smith}}]{Huber_2011}
{Huber}, D., {Bedding}, T.~R., {Stello}, D., {et~al.} 2011, \apj, 743, 143, \dodoi{10.1088/0004-637X/743/2/143}

\bibitem[{Huber {et~al.}(2023)Huber, Pinsonneault, Beck, Bedding, Bland-Hawthorn, Breton, Bugnet, Chaplin, Garcia, Grunblatt, Guzik, Hekker, Kawaler, Mathis, Mathur, Metcalfe, Mosser, Ness, Piro, Serenelli, Sharma, Soderblom, Stassun, Stello, Tayar, van Belle, \& Zinn}]{Huber_2023_Roman}
Huber, D., Pinsonneault, M., Beck, P., {et~al.} 2023, Asteroseismology with the Roman Galactic Bulge Time-Domain Survey.
\newblock \doarXiv{2307.03237}

\bibitem[{{Kallinger} {et~al.}(2010){Kallinger}, {Mosser}, {Hekker}, {Huber}, {Stello}, {Mathur}, {Basu}, {Bedding}, {Chaplin}, {De Ridder}, {Elsworth}, {Frandsen}, {Garc{\'\i}a}, {Gruberbauer}, {Matthews}, {Borucki}, {Bruntt}, {Christensen-Dalsgaard}, {Gilliland}, {Kjeldsen}, \& {Koch}}]{Kallinger_2010_CAN}
{Kallinger}, T., {Mosser}, B., {Hekker}, S., {et~al.} 2010, \aap, 522, A1, \dodoi{10.1051/0004-6361/201015263}

\bibitem[{{Lightkurve Collaboration} {et~al.}(2018){Lightkurve Collaboration}, {Cardoso}, {Hedges}, {Gully-Santiago}, {Saunders}, {Cody}, {Barclay}, {Hall}, {Sagear}, {Turtelboom}, {Zhang}, {Tzanidakis}, {Mighell}, {Coughlin}, {Bell}, {Berta-Thompson}, {Williams}, {Dotson}, \& {Barentsen}}]{Lightkurve_2018}
{Lightkurve Collaboration}, {Cardoso}, J.~V.~d.~M., {Hedges}, C., {et~al.} 2018, {Lightkurve: Kepler and TESS time series analysis in Python}, Astrophysics Source Code Library.
\newblock \doeprint{1812.013}

\bibitem[{{Lomb}(1976)}]{Lomb_1976}
{Lomb}, N.~R. 1976, \apss, 39, 447, \dodoi{10.1007/BF00648343}

\bibitem[{{Mackereth} {et~al.}(2021){Mackereth}, {Miglio}, {Elsworth}, {Mosser}, {Mathur}, {Garcia}, {Nardiello}, {Hall}, {Vrard}, {Ball}, {Basu}, {Beaton}, {Beck}, {Bergemann}, {Bossini}, {Casagrande}, {Campante}, {Chaplin}, {Chiappini}, {Girardi}, {J{\o}rgensen}, {Khan}, {Montalb{\'a}n}, {Nielsen}, {Pinsonneault}, {Rodrigues}, {Serenelli}, {Silva Aguirre}, {Stello}, {Tayar}, {Teske}, {van Saders}, \& {Willett}}]{Mackereth_2021MNRAS.502.1947M}
{Mackereth}, J.~T., {Miglio}, A., {Elsworth}, Y., {et~al.} 2021, \mnras, 502, 1947, \dodoi{10.1093/mnras/stab098}

\bibitem[{{Mathur} {et~al.}(2010){Mathur}, {Garc{\'\i}a}, {R{\'e}gulo}, {Creevey}, {Ballot}, {Salabert}, {Arentoft}, {Quirion}, {Chaplin}, \& {Kjeldsen}}]{Mathur_2010_A2Z}
{Mathur}, S., {Garc{\'\i}a}, R.~A., {R{\'e}gulo}, C., {et~al.} 2010, \aap, 511, A46, \dodoi{10.1051/0004-6361/200913266}

\bibitem[{{Miglio} {et~al.}(2021){Miglio}, {Chiappini}, {Mackereth}, {Davies}, {Brogaard}, {Casagrande}, {Chaplin}, {Girardi}, {Kawata}, {Khan}, {Izzard}, {Montalb{\'a}n}, {Mosser}, {Vincenzo}, {Bossini}, {Noels}, {Rodrigues}, {Valentini}, \& {Mandel}}]{Miglio_2021}
{Miglio}, A., {Chiappini}, C., {Mackereth}, J.~T., {et~al.} 2021, \aap, 645, A85, \dodoi{10.1051/0004-6361/202038307}

\bibitem[{{Mosser} \& {Appourchaux}(2009)}]{Mosser_2009_COR}
{Mosser}, B., \& {Appourchaux}, T. 2009, \aap, 508, 877, \dodoi{10.1051/0004-6361/200912944}

\bibitem[{Mosser {et~al.}(2018)Mosser, Gehan, Belkacem, Samadi, Michel, \& Goupil}]{Mosser_2018}
Mosser, B., Gehan, C., Belkacem, K., {et~al.} 2018, \aap, 618, A109, \dodoi{10.1051/0004-6361/201832777}

\bibitem[{{Mosser} {et~al.}(2015){Mosser}, {Vrard}, {Belkacem}, {Deheuvels}, \& {Goupil}}]{Mosser_2015}
{Mosser}, B., {Vrard}, M., {Belkacem}, K., {Deheuvels}, S., \& {Goupil}, M.~J. 2015, \aap, 584, A50, \dodoi{10.1051/0004-6361/201527075}

\bibitem[{{Mosser} {et~al.}(2012{\natexlab{a}}){Mosser}, {Goupil}, {Belkacem}, {Michel}, {Stello}, {Marques}, {Elsworth}, {Barban}, {Beck}, {Bedding}, {De Ridder}, {Garc{\'\i}a}, {Hekker}, {Kallinger}, {Samadi}, {Stumpe}, {Barclay}, \& {Burke}}]{Mosser_2012_evolution}
{Mosser}, B., {Goupil}, M.~J., {Belkacem}, K., {et~al.} 2012{\natexlab{a}}, \aap, 540, A143, \dodoi{10.1051/0004-6361/201118519}

\bibitem[{{Mosser} {et~al.}(2012{\natexlab{b}}){Mosser}, {Elsworth}, {Hekker}, {Huber}, {Kallinger}, {Mathur}, {Belkacem}, {Goupil}, {Samadi}, {Barban}, {Bedding}, {Chaplin}, {Garc{\'\i}a}, {Stello}, {De Ridder}, {Middour}, {Morris}, \& {Quintana}}]{Mosser_2012_power_excess}
{Mosser}, B., {Elsworth}, Y., {Hekker}, S., {et~al.} 2012{\natexlab{b}}, \aap, 537, A30, \dodoi{10.1051/0004-6361/201117352}

\bibitem[{{Pires} {et~al.}(2015){Pires}, {Mathur}, {Garc{\'\i}a}, {Ballot}, {Stello}, \& {Sato}}]{Pires_2015}
{Pires}, S., {Mathur}, S., {Garc{\'\i}a}, R.~A., {et~al.} 2015, \aap, 574, A18, \dodoi{10.1051/0004-6361/201322361}

\bibitem[{{Rauer} {et~al.}(2024){Rauer}, {Aerts}, {Cabrera}, {Deleuil}, {Erikson}, {Gizon}, {Goupil}, {Heras}, {Lorenzo-Alvarez}, {Marliani}, {Martin-Garcia}, {Mas-Hesse}, {O'Rourke}, {Osborn}, {Pagano}, {Piotto}, {Pollacco}, {Ragazzoni}, {Ramsay}, {Udry}, {Appourchaux}, {Benz}, {Brandeker}, {G{\"u}del}, {Janot-Pacheco}, {Kabath}, {Kjeldsen}, {Min}, {Santos}, {Smith}, {Suarez}, {Werner}, {Aboudan}, {Abreu}, {Acu{\~n}a}, {Adams}, {Adibekyan}, {Affer}, {Agneray}, {Agnor}, {Aguirre B{\o}rsen-Koch}, {Ahmed}, {Aigrain}, {Al-Bahlawan}, {Alcacera Gil}, {Alei}, {Alencar}, {Alexander}, {Alfonso-Garz{\'o}n}, {Alibert}, {Allende Prieto}, {Almeida}, {Alonso Sobrino}, {Altavilla}, {Althaus}, {Alonso Alvarez Trujillo}, {Amarsi}, {Ammler-von Eiff}, {Am{\^o}res}, {Andrade}, {Antoniadis-Karnavas}, {Ant{\'o}nio}, {Aparicio del Moral}, {Appolloni}, {Arena}, {Armstrong}, {Aroca Aliaga}, {Asplund}, {Audenaert}, {Auricchio}, {Avelino}, {Baeke}, {Bailli{\'e}}, {Balado}, {Balestra}, {Ball}, {Ballans}, {Ballot}, {Barban}, {Barbary}, {Barbieri}, {Barcel{\'o} Forteza}, {Barker}, {Barklem}, {Barnes}, {Barrado Navascues}, {Barragan}, {Baruteau}, {Basu}, {Baudin}, {Baumeister}, {Bayliss}, {Bazot}, {Beck}, {Bedding}, {Belkacem}, {Bellinger}, {Benatti}, {Benomar}, {B{\'e}rard}, {Bergemann}, {Bergomi}, {Bernardo}, {Biazzo}, {Bignamini}, {Bigot}, {Billot}, {Binet}, {Biondi}, {Biondi}, {Birch}, {Bitsch}, {Bluhm Ceballos}, {B{\'o}di}, {Bogn{\'a}r}, {Boisse}, {Bolmont}, {Bonanno}, {Bonavita}, {Bonfanti}, {Bonfils}, {Bonito}, {Bonomo}, {B{\"o}rner}, {Boro Saikia}, {Borreguero Mart{\'\i}n}, {Borsa}, {Borsato}, {Bossini}, {Bouchy}, {Bou{\'e}}, {Boufleur}, {Boumier}, {Bourrier}, {Bowman}, {Bozzo}, {Bradley}, {Bray}, {Bressan}, {Breton}, {Brienza}, {Brito}, {Brogi}, {Brown}, {Brown}, {Brun}, {Bruno}, {Bruns}, {Buchhave}, {Bugnet}, {Buldgen}, {Burgess}, {Busatta}, {Busso}, {Buzasi}, {Caballero}, {Cabral}, {Calderone}, {Cameron}, {Cameron}, {Campante}, {Canto Martins}, {Cara}, {Carone}, {Carrasco}, {Casagrande}, {Casewell}, {Cassisi}, {Castellani}, {Castro}, {Catala}, {Catal{\'a}n Fern{\'a}ndez}, {Catelan}, {Cegla}, {Cerruti}, {Cessa}, {Chadid}, {Chaplin}, {Charpinet}, {Chiappini}, {Chiarucci}, {Chiavassa}, {Chinellato}, {Chirulli}, {Christensen-Dalsgaard}, {Church}, {Claret}, {Clarke}, {Claudi}, {Clermont}, {Coelho}, {Coelho}, {Cogato}, {Colom{\'e}}, {Condamin}, {Conseil}, {Corbard}, {Correia}, {Corsaro}, {Cosentino}, {Costes}, {Cottinelli}, {Covone}, {Creevey}, {Crida}, {Csizmadia}, {Cunha}, {Curry}, {da Costa}, {da Silva}, {Dalal}, {Damasso}, {Damiani}, {Damiani}, {Liduina das Chagas}, {Davies}, {Davies}, {Davies}, {Davison}, {de Almeida}, {de Angeli}, {Cabral de Barros}, {de Castro Le{\~a}o}, {Brito de Freitas}, {de Freitas}, {De Martino}, {Renan de Medeiros}, {de Paula}, {de Plaa}, {De Ridder}, {Deal}, {Decin}, {Deeg}, {Degl'Innocenti}, {Deheuvels}, {del Burgo}, {Del Sordo}, {Delgado-Mena}, {Demangeon}, {Denk}, {Derekas}, {Desidera}, {Dexet}, {Di Criscienzo}, {Di Giorgio}, {Di Mauro}, {Diaz Rial}, {D{\'\i}az-Garc{\'\i}a}, {Dima}, {Dinuzzi}, {Dionatos}, {Distefano}, {do Nascimento}, {Domingo}, {D'Orazi}, {Dorn}, {Doyle}, {Duarte}, {Ducellier}, {Dumaye}, {Dumusque}, {Dupret}, {Eggenberger}, {Ehrenreich}, {Eigm{\"u}ller}, {Eising}, {Emilio}, {Eriksson}, {Ermocida}, {Isidoro Escate Giribaldi}, {Eschen}, {Estrela}, {Evans}, {Fabbian}, {Fabrizio}, {Faria}, {Farina}, {Farinato}, {Feliz}, {Feltzing}, {Fenouillet}, {Ferrari}, {Ferraz-Mello}, {Fialho}, {Fienga}, {Figueira}, {Fiori}, {Flaccomio}, {Focardi}, {Foley}, {Fontignie}, {Ford}, {Fornazier}, {Forveille}, {Fossati}, {de Marca Franca}, {da Silva}, {Frasca}, {Fridlund}, {Furlan}, {Gabler}, {Gaido}, {Gallagher}, {Galli}, {Garcia}, {Garc{\'\i}a Hern{\'a}ndez}, {Garcia Munoz}, {Garc{\'\i}a-V{\'a}zquez}, {Garrido Haba}, {Gaulme}, {Gauthier}, {Gehan}, {Gent}, {Georgieva}, {Ghigo}, {Giana}, {Gill}, {Girardi}, {Giuliatti Winter}, {Giusi}, {Gomes da Silva}, {G{\'o}mez Zazo}, {Gomez-Lopez}, {Isai Gonz{\'a}lez Hern{\'a}ndez}, {Gonzalez Murillo}, {Gorius}, {Gouel}, {Goulty}, {Granata}, {Grenfell}, {Grie{\ss}bach}, {Grolleau}, {Grouffal}, {Grziwa}, {Guarcello}, {Gueguen}, {Guenther}, {Guilhem}, {Guillerot}, {Guiot}, {Guterman}, {Guti{\'e}rrez}, {Guti{\'e}rrez-Canales}, {Hagelberg}, {Haldemann}, {Hall}, {Handberg}, {Harrison}, {Harrison}, {Hasiba}, {Haswell}, {Hatalova}, {Hatzes}, {Haywood}, {H{\'e}brard}, {Heckes}, {Heiter}, {Hekker}, {Heller}, {Helling}, {Helminiak}, {Hemsley}, {Heng}, {Hermans}, {Hermes}, {Hidalgo Torres}, {Hinkel}, {Hobbs}, {Hodgkin}, {Hofmann}, {Hojjatpanah}, {Houdek}, {Huber}, {Huesler}, {Hui-Bon-Hoa}, {Huygen}, {Huynh}, {Iro}, {Irwin}, {Irwin}, {Izidoro}, {Jacquinod}, {Emborg Jannsen}, {Janson}, {Jeszenszky}, {Jiang}, {Jos{\'e} Jimenez Mancebo}, {Jofre}, {Johansen}, {Johnston}, {Jones}, {Kallinger}, {K{\'a}lm{\'a}n}, {Kanitz}, {Karjalainen}, {Karjalainen}, {Karoff}, {Kawaler}, {Kawata}, {Keereman}, {Keiderling}, {Kennedy}, {Kenworthy}, {Kerschbaum}, {Kidger}, {Kiefer}, {Kintziger}, {Kislyakova}, {Kiss}, {Klagyivik}, {Klahr}, {Klevas}, {Kochukhov}, {K{\"o}hler}, {Kolb}, {Koncz}, {Korth}, {Kostogryz}, {Kov{\'a}cs}, {Kov{\'a}cs}, {Kozhura}, {Krivova}, {Ku{\v{c}}inskas}, {Kuhlemann}, {Kupka}, {Laauwen}, {Labiano}, {Lagarde}, {Laget}, {Laky}, {Lam}, {Lambrechts}, {Lammer}, {Lanza}, {Lanzafame}, {Lares Martiz}, {Laskar}, {Latter}, {Lavanant}, {Lawrenson}, {Lazzoni}, {Lebre}, {Lebreton}, {Lecavelier des Etangs}, {Leinhardt}, {Leleu}, {Lendl}, {Leto}, {Levillain}, {Libert}, {Lichtenberg}, {Ligi}, {Lignieres}, {Lillo-Box}, {Linsky}, {Scige Liu}, {Loidolt}, {Longval}, {Lopes}, {Lorenzani}, {Ludwig}, {Lund}, {Sloth Lundkvist}, {Luri}, {Maceroni}, {Madden}, {Madhusudhan}, {Maggio}, {Magliano}, {Magrin}, {Mahy}, {Maibaum}, {Malac-Allain}, {Malapert}, {Malavolta}, {Maldonado}, {Mamonova}, {Manchon}, {Mann}, {Mantovan}, {Marafatto}, {Marconi}, {Mardling}, {Marigo}, {Marinoni}, {Marques}, {Marques}, {Marrese}, {Marshall}, {Mart{\'\i}nez Perales}, {Mary}, {Marzari}, {Masana}, {Mascher}, {Mathis}, {Mathur}, {Mattiuci Figueiredo}, {Maxted}, {Mazeh}, {Mazevet}, {Mazzei}, {McCormac}, {McMillan}, {Menou}, {Merle}, {Meru}, {Mesa}, {Messina}, {M{\'e}sz{\'a}ros}, {Meunier}, {Meunier}, {Micela}, {Michaelis}, {Michel}, {Michielsen}, {Michtchenko}, {Miglio}, {Miguel}, {Milligan}, {Mirouh}, {Mitchell}, {Moedas}, {Molendini}, {Moln{\'a}r}, {Mombarg}, {Montalban}, {Montalto}, {Monteiro}, {Morales}, {Morales-Calderon}, {Morbidelli}, {Mordasini}, {Moreau}, {Morel}, {Morello}, {Morin}, {Mortier}, {Mosser}, {Mourard}, {Mousis}, {Moutou}, {Mowlavi}, {Moya}, {Muehlmann}, {Muirhead}, {Munari}, {Musella}, {Mustill}, {Nardetto}, {Nardiello}, {Narita}, {Nascimbeni}, {Nash}, {Neiner}, {Nelson}, {Nettelmann}, {Nicolini}, {Nielsen}, {Niemi}, {Noack}, {Noels-Grotsch}, {Noll}, {Norazman}, {Norton}, {Nsamba}, {Ofir}, {Ogilvie}, {Olander}, {Olivetto}, {Olofsson}, {Ong}, {Ortolani}, {Oshagh}, {Ottacher}, {Ottensamer}, {Ouazzani}, {Paardekooper}, {Pace}, {Pajas}, {Palacios}, {Palandri}, {Palle}, {Paproth}, {Parro}, {Parviainen}, {Granado}, {Passegger}, {Pastor-Morales}, {P{\"a}tzold}, {Gade Pedersen}, {Pena Hidalgo}, {Pepe}, {Pereira}, {Persson}, {Pertenais}, {Peter}, {Petit}, {Petit}, {Pezzuto}, {Pichierri}, {Pietrinferni}, {Pinheiro}, {Pinsonneault}, {Plachy}, {Plasson}, {Plez}, {Poppenhaeger}, {Poretti}, {Portaluri}, {Portell}, {Frederico Porto de Mello}, {Poyatos}, {Pozuelos}, {Prada Moroni}, {Pricopi}, {Prisinzano}, {Quade}, {Quirrenbach160}, {Rabanal Reina6}, {Rabello Soares}, {Raimondo}, {Rainer}, {Ram{\'o}n Rod{\'o}n}, {Ram{\'o}n-Ballesta}, {Ramos Zapata}, {R{\"a}tz}, {Rauterberg}, {Redman}, {Redmer}, {Reese}, {Regibo}, {Reiners}, {Reinhold}, {Renie}, {Ribas}, {Ribeiro}, {Pereira Ricciardi}, {Rice}, {Richard}, {Riello}, {Rieutord}, {Ripepi}, {Rixon}, {Rockstein}, {Rodr{\'\i}guez}, {Rodr{\'\i}guez D{\'\i}az}, {Rodriguez Garcia}, {Rodriguez-Gomez}, {Roehlly}, {Roig}, {Rojas-Ayala}, {Rolf}, {Lysgaard R{\o}rsted}, {Rosado}, {Rosotti}, {Roth}, {Roth}, {Rousseau}, {Roxburgh}, {Roy}, {Royer}, {Ruane}, {Rufini Mastropasqua}, {Ruiz de Galarreta}, {Russi}, {Saar}, {Saillenfest}, {Salaris}, {Salmon}, {Saltas}, {Samadi}, {Samadi}, {Samra}, {Sanches da Silva}, {Andr{\'e}s S{\'a}nchez Carrasco}, {Santerne}, {Santoli}, {Santos}, {Sanz Mesa}, {Sarro}, {Scandariato}, {Sch{\"a}fer}, {Schlafly}, {Schmider}, {Schneider}, {Schou}, {Schunker}, {J{\"o}rg Schwarzkopf}, {Serenelli}, {Seynaeve}, {Shan}, {Shapiro}, {Shipman}, {Sicilia}, {Sierra Sanmartin}, {Sigot}, {Silliman}, {Silvotti}, {Simon}, {Simoyama Napoli}, {Skarka}, {Smalley}, {Smiljanic}, {Smit}, {Smith}, {Smith}, {Snellen}, {S{\'o}dor}, {Sohl}, {Solanki}, {Sortino}, {Sousa}, {Southworth}, {Souto}, {Sozzetti}, {Stamatellos}, {Stassun}, {Steller}, {Stello}, {Stelzer}, {Stiebeler}, {Stokholm}, {Storelvmo}, {Strassmeier}, {Str{\o}m}, {Strugarek}, {Sulis}, {{\v{S}}vanda}, {Szabados}, {Szab{\'o}}, {Szab{\'o}}, {Szuszkiewicz}, {Talens}, {Teti}, {Theisen}, {Th{\'e}venin}, {Thoul}, {Tiphene}, {Titz-Weider}, {Tkachenko}, {Tomecki}, {Tonfat}, {Tosi}, {Trampedach}, {Traven}, {Triaud}, {Tr{\o}nnes}, {Tsantaki}, {Tschentscher}, {Turin}, {Tvaruzka}, {Ulmer}, {Ulmer-Moll}, {Ulusoy}, {Umbriaco}, {Valencia}, {Valentini}, {Valio}, {Valverde Guijarro}, {Van Eylen}, {Van Grootel}, {van Kempen}, {Van Reeth}, {Van Zelst}, {Vandenbussche}, {Vasiliou}, {Vasilyev}, {Vaz de Mascarenhas}, {Vazan}, {Vela Nunez}, {Nunes Velloso}, {Ventura}, {Ventura}, {Venturini}, {Trallero}, {Veras}, {Verdugo}, {Verma}, {Vibert}, {Vicanek Martinez}, {Vida}, {Vigan}, {Villacorta}, {Villaver}, {Villaverde Aparicio}, {Viotto}, {Vorobyov}, {Vorontsov}, {Wagner}, {Walloschek}, {Walton}, {Walton}, {Wang}, {Waters}, {Watson}, {Wedemeyer}, {Weeks}, {Weingril}, {Weiss}, {Wendler}, {West}, {Westerdorff}, {Westphal}, {Wheatley}, {White}, {Whittaker}, {Wickhusen}, {Wilson}, {Windsor}, {Winter}, {Lykke Winther}, {Winton}, {Witteck}, {Witzke}, {Woitke}, {Wolter}, {Wuchterl}, {Wyatt}, {Yang}, {Yu}, {Zanmar Sanchez}, {Rosa Zapatero Osorio}, {Zechmeister},
  {Zhou}, {Ziemke}, \& {Zwintz}}]{plato_mission_2024}
{Rauer}, H., {Aerts}, C., {Cabrera}, J., {et~al.} 2024, arXiv e-prints, arXiv:2406.05447, \dodoi{10.48550/arXiv.2406.05447}

\bibitem[{Reback {et~al.}(2022)Reback, jbrockmendel, McKinney, den Bossche, Roeschke, Augspurger, Hawkins, Cloud, gfyoung, Sinhrks, Hoefler, Klein, Petersen, Tratner, She, Ayd, Naveh, Darbyshire, Shadrach, Garcia, Schendel, Hayden, Saxton, Gorelli, Li, Wörtwein, Zeitlin, Jancauskas, McMaster, \& Li}]{reback_2022_6702671}
Reback, J., jbrockmendel, McKinney, W., {et~al.} 2022, pandas-dev/pandas: Pandas 1.4.3, v1.4.3,  Zenodo, \dodoi{10.5281/zenodo.6702671}

\bibitem[{Reyes {et~al.}(2022)Reyes, Stello, Hon, \& Zinn}]{reyes2022VettingAsteroseismicDn}
Reyes, C., Stello, D., Hon, M., \& Zinn, J.~C. 2022, Monthly Notices of the Royal Astronomical Society, 511, 5578, \dodoi{10.1093/mnras/stac445}

\bibitem[{Richard \& Lippmann(1991)}]{richard1991_nn_bayesian_dist}
Richard, M.~D., \& Lippmann, R.~P. 1991, Neural Computation, 3, 461, \dodoi{10.1162/neco.1991.3.4.461}

\bibitem[{{Ricker} {et~al.}(2015){Ricker}, {Winn}, {Vanderspek}, {Latham}, {Bakos}, {Bean}, {Berta-Thompson}, {Brown}, {Buchhave}, {Butler}, {Butler}, {Chaplin}, {Charbonneau}, {Christensen-Dalsgaard}, {Clampin}, {Deming}, {Doty}, {De Lee}, {Dressing}, {Dunham}, {Endl}, {Fressin}, {Ge}, {Henning}, {Holman}, {Howard}, {Ida}, {Jenkins}, {Jernigan}, {Johnson}, {Kaltenegger}, {Kawai}, {Kjeldsen}, {Laughlin}, {Levine}, {Lin}, {Lissauer}, {MacQueen}, {Marcy}, {McCullough}, {Morton}, {Narita}, {Paegert}, {Palle}, {Pepe}, {Pepper}, {Quirrenbach}, {Rinehart}, {Sasselov}, {Sato}, {Seager}, {Sozzetti}, {Stassun}, {Sullivan}, {Szentgyorgyi}, {Torres}, {Udry}, \& {Villasenor}}]{Tess_mission_ricker_2015}
{Ricker}, G.~R., {Winn}, J.~N., {Vanderspek}, R., {et~al.} 2015, Journal of Astronomical Telescopes, Instruments, and Systems, 1, 014003, \dodoi{10.1117/1.JATIS.1.1.014003}

\bibitem[{Rothe {et~al.}(2015)Rothe, Timofte, \& Van~Gool}]{Rothe_2015}
Rothe, R., Timofte, R., \& Van~Gool, L. 2015, in 2015 IEEE International Conference on Computer Vision Workshop (ICCVW), 252--257, \dodoi{10.1109/ICCVW.2015.41}

\bibitem[{{Scargle}(1982)}]{Scargle_1982}
{Scargle}, J.~D. 1982, \apj, 263, 835, \dodoi{10.1086/160554}

\bibitem[{{Sharma} {et~al.}(2019){Sharma}, {Stello}, {Bland-Hawthorn}, {Hayden}, {Zinn}, {Kallinger}, {Hon}, {Asplund}, {Buder}, {De Silva}, {D'Orazi}, {Freeman}, {Kos}, {Lewis}, {Lin}, {Lind}, {Martell}, {Simpson}, {Wittenmyer}, {Zucker}, {Zwitter}, {Bedding}, {Chen}, {Cotar}, {Esdaile}, {Horner}, {Huber}, {Kafle}, {Khanna}, {Li}, {Ting}, {Nataf}, {Nordlander}, {Saadon}, {Traven}, {Wright}, \& {Wyse}}]{Sharma_2019}
{Sharma}, S., {Stello}, D., {Bland-Hawthorn}, J., {et~al.} 2019, \mnras, 490, 5335, \dodoi{10.1093/mnras/stz2861}

\bibitem[{Spergel {et~al.}(2015)Spergel, Gehrels, Baltay, Bennett, Breckinridge, Donahue, Dressler, Gaudi, Greene, Guyon, Hirata, Kalirai, Kasdin, Macintosh, Moos, Perlmutter, Postman, Rauscher, Rhodes, Wang, Weinberg, Benford, Hudson, Jeong, Mellier, Traub, Yamada, Capak, Colbert, Masters, Penny, Savransky, Stern, Zimmerman, Barry, Bartusek, Carpenter, Cheng, Content, Dekens, Demers, Grady, Jackson, Kuan, Kruk, Melton, Nemati, Parvin, Poberezhskiy, Peddie, Ruffa, Wallace, Whipple, Wollack, \& Zhao}]{Spergel_2015_Roman}
Spergel, D., Gehrels, N., Baltay, C., {et~al.} 2015, Wide-Field InfrarRed Survey Telescope-Astrophysics Focused Telescope Assets WFIRST-AFTA 2015 Report.
\newblock \doarXiv{1503.03757}

\bibitem[{{Sreenivas} {et~al.}(2026){Sreenivas}, {Bedding}, {Huber}, {Stello}, {Hon}, {Reyes}, {Li}, \& {Hey}}]{Sreenivas_2026}
{Sreenivas}, K.~R., {Bedding}, T.~R., {Huber}, D., {et~al.} 2026, \mnras, 548, stag671, \dodoi{10.1093/mnras/stag671}

\bibitem[{{Stello} {et~al.}(2009){Stello}, {Chaplin}, {Basu}, {Elsworth}, \& {Bedding}}]{Stello_2009}
{Stello}, D., {Chaplin}, W.~J., {Basu}, S., {Elsworth}, Y., \& {Bedding}, T.~R. 2009, \mnras, 400, L80, \dodoi{10.1111/j.1745-3933.2009.00767.x}

\bibitem[{{Stello} {et~al.}(2022){Stello}, {Saunders}, {Grunblatt}, {Hon}, {Reyes}, {Huber}, {Bedding}, {Elsworth}, {Garc{\'\i}a}, {Hekker}, {Kallinger}, {Mathur}, {Mosser}, \& {Pinsonneault}}]{Stello_2022MNRAS.512.1677S}
{Stello}, D., {Saunders}, N., {Grunblatt}, S., {et~al.} 2022, \mnras, 512, 1677, \dodoi{10.1093/mnras/stac414}

\bibitem[{{Stewart} {et~al.}(2022){Stewart}, {Bach}, {Berthet}, \& {Vert}}]{Stewart_2022arXiv221105641S}
{Stewart}, L., {Bach}, F., {Berthet}, Q., \& {Vert}, J.-P. 2022, arXiv e-prints, arXiv:2211.05641, \dodoi{10.48550/arXiv.2211.05641}

\bibitem[{{Yu} {et~al.}(2018){Yu}, {Huber}, {Bedding}, {Stello}, {Hon}, {Murphy}, \& {Khanna}}]{Yu_2018}
{Yu}, J., {Huber}, D., {Bedding}, T.~R., {et~al.} 2018, \apjs, 236, 42, \dodoi{10.3847/1538-4365/aaaf74}

\end{thebibliography}
\bibliographystyle{aasjournal}

\end{document}